\documentclass[%
 preprint,
superscriptaddress,
 amsmath,amssymb,
 aps,
]{revtex4-2}

\usepackage{graphicx}
\usepackage{dcolumn}
\usepackage{bm}
\usepackage{color,soul}

\usepackage{epstopdf}

\newcommand{\be}{\begin{eqnarray}}
\newcommand{\ee}{\end{eqnarray}}

\begin{document}

\title{Quantum-critical transport in marginal Fermi liquids}
\author{Hideaki Maebashi}
\affiliation{Department of Physics, University of Tokyo, Tokyo 113-0033, Japan}

\author{Chandra M. Varma}
\affiliation{\footnote{Recalled Professor}Department of Physics, University of California, Berkeley, CA. 94704, USA}
\affiliation{\footnote{Emeritus Distinguished Professor}Department of Physics, University of California, Riverside, CA. 92521, USA}
\date{\today}

\begin{abstract}
We use the Kubo response functions to calculate the electrical and thermal conductivity and Seebeck coefficient at low temperatures and frequencies in the quantum-critical region for fermions on a lattice. The theory uses scattering of the fermions with the previously derived collective fluctuations caused by topological defects of the quantum XY model coupled to fermions. As in the marginal Fermi-liquid phenomenology, the fluctuations are a scale invariant function ${\cal{F}}(\omega/T)$, but unlike it, they have both a momentum dependence and a momentum-dependent coupling to fermions. They have the unusual form of being products of ${\cal{F}}(\omega/T)$ and a function of the momentum transfer. This gives that there is no vertex correction to the single-particle self-energy over a range of momenta and frequencies near the Fermi-surface, unlike Migdal's theorem. The imaginary part of the single-particle self-energy then continues to be linear in max$(\omega,T)$ with a weak  momentum dependence for a range of momentum near the Fermi surface, as in the phenomenology. This is used to solve the vertex equation in the Kubo formula for the transport properties at low temperatures with multiplicative corrections of $O(T/E_F)$. The microscopic model is applicable to the fluctuations of the loop-current order in cuprates as well as to a class of quasi-two-dimensional heavy-fermion and other metallic antiferromagnets, and was proposed recently also for the possible loop-current order in Moir\'{e} twisted bilayer graphene and bilayer WSe$_2$. All these metals have a linear-in-temperature electrical resistivity in the quantum-critical region of their phase diagrams, often termed ``Planckian" resistivity. The solution of the  Kubo equation for transport shows that vertex renormalizations to the external fields, besides those caused by Aslamazov-Larkin (A-L) processes, are absent. A-L appears as an Umklapp scattering matrix, which gives a temperature-independent multiplicative factor for the electrical resistivity but does not affect the thermal conductivity. We also show that the mass renormalization which gives a logarithmic enhancement of the marginal Fermi-liquid specific heat does not appear in the electrical resistivity and, more remarkably, in the thermal conductivity. On the other hand the mass renormalization $\propto \ln \omega_c/T$ ($\omega_c$ is the upper cutoff of the fluctuations), appears in  the Seebeck coefficient. We also discuss in detail the conservation laws that play a crucial role in all transport properties. We  calculate exactly, the numerical coefficients of the transport properties for a circular Fermi surface. The leading temperature dependence is shown to remain the same for a general Fermi surface, but it is too messy to calculate the numerical coefficient. 
\end{abstract}
\maketitle
\newpage
\noindent

\section{Introduction} 

The discovery of normal state properties in cuprates that violate the quasiparticle concept of Landau Fermi liquids have been intensely discussed in the last three decades \cite{Ginzburg-rev}. The most prominently discussed  \cite{Varma-rmp2020} of these properties is the resistivity which has a linear in temperature dependence from the lowest temperature investigated by suppressing superconductivity  often up to temperatures of $O(10^3)$ K. Optical conductivity, Raman scattering, nuclear relaxation rates, tunneling conductance, etc. were similarly discovered to have anomalous but simple frequency and temperature dependence. Soon thereafter, similar properties were discovered in heavy fermion compounds, for reviews see \cite{HvLRMP2007, PaschenYBRh}, and in the normal state of high temperature Fe-based compounds, for reviews see \cite{ShibauchiQCP, Hosono-rev}, and more recently in twisted bi-layer graphene (TBG) \cite{Herrero2019, Efetov2019} and twisted bi-layer WSe$_2$ \cite{Pasupathy2021, MakShanWSe2} (TBWSe).

All the anomalous properties in all these metals in their critical regime follow from the marginal Fermi-liquid (MFL) phenomenological hypothesis \cite{CMV-MFL} that there must exist fluctuations whose absorptive part is a scale invariant function ${\cal F}(\omega/T)$, equivalently  $\propto 1/\tau$ where $\tau$ is the imaginary time rather than $1/\tau^2$ as in a Fermi liquid, and only weakly dependent on momentum. Several predictions of this hypothesis were experimentally verified. Especially important for this paper is the prediction, verified with angle resolved photoemission (ARPES) \cite{Johnson2000, KaminskiPR2005, Bok_ScienceADV, AbrahamsV-PNAS} and angular dependence of magnetoresistance \cite{Ramshaw2021} that the imaginary part of the single-particle self-energy is proportional to ${\rm max}(\omega, T)$ and only weakly dependent on momentum.  It had been asserted that the transport scattering rate for electrical and thermal conductivity have the same frequency and temperature dependence as the single-particle scattering rate because the fluctuations had been assumed momentum independent. In this paper this issue and the ratio between the two scattering rates is investigated precisely, based on derived fluctuations that are both momentum dependent and have momentum-dependent interaction with fermions. We also derive the Seebeck coefficient and show that it has the $T \ln \frac{\omega_c}{T}$ form as verified in experiments \cite{Seebeck_LaNdSrCuO,Seebeck_Bi2201_PhysRevB.104.014515,Seebeck_cuprates,Seebeck_Fe_pnictides_PhysRevB.79.104504}. This logarithmic enhancement is inherited from the mass enhancement which appears in the specific heat near the quantum-critical point. $\omega_c$ is the upper cutoff in the fluctuation spectra which may therefore be obtained from the measured thermopower. Contrary to what one may naively think, neither the thermal conductivity nor the electrical conductivity have a mass enhancement. This is already verified in experiments because they show a pure linear-in-$T$ resistivity in addition to a constant caused by impurity scattering.

MFL posited a singularity at $T = 0$, i.e. a quantum-critical point \cite{VNS}.  It was suggested that there must be a phase transition ending at a quantum-critical point  as a function of doping \cite{CMVlosalamos1991} in cuprates. Its  physical nature as a loop-current order, odd in time-reversal and inversion was predicted \cite{cmv1997, simon-cmv} using a microscopic model taking into account the charge transfer nature of the cuprates \cite{VSA1987}. The model for the quantum-critical fluctuations of  the order parameter is the quantum XY-model coupled to fermions (QXY-F model) \cite{Aji-V-qcf1}. The applicability of the QXY-F model to antiferromagnetic quantum-critical points in heavy fermions and Fe-based compounds has been shown \cite{cmv_PhysRevLett.115.186405,CMV-IOP-REV}, as also for TBG and TBWSe \cite{ChubukovV2025}. The fluctuations of this model have been derived analytically \cite{Hou-CMV-RG} as well as through quantum Monte-Carlo calculations \cite{ZhuChenCMV2015, ZhuHouV2016} and are functions of $\omega/T$ as in the MFL hypothesis but the momentum dependence of the fluctuations and the vertex to fermions have important differences, which conspire to give similar but not identical results in the microscopic theory and that derived from the phenomenological hypothesis \cite{CMV-MFL}.  Direct evidence of the fluctuations of the model over the entire momentum region is found by neutron scattering in heavy fermions and in Fe-based antiferromagnets \cite{CMVZhuSchroeder2015, Schroeder1998, Schroeder2000}. The fluctuations of the loop-current order in cuprates are spread out over a range of about 0.4 eV so that even though their integrated value is about $(0.1 \mu_B)^2$ per unit cell, they are very hard to discover by inelastic neutron scattering. However, their manifestation in the calculated density correlations \cite{Shekhter-V-Hydro, CMV_Sqw2017} has been directly observed through measurements of the dynamic structure factor \cite{Abbamonte2018} and earlier for long wavelength excitations by Raman scattering \cite{Klein1991}.
 
A variety of experiments are consistent \cite{Fauque2006, Greven2011, Kaminski-diARPES, Hsieh2017, Armitage-Biref, CMVEurophys2014, Shu2018, Kapitulnik1, cmv-kerr, Kapitulnik-galmag, Bourges2021}
 with the predicted broken symmetry in cuprates, while the broken symmetry relevant to the quantum fluctuations in the heavy fermions and other metallic antiferromagnets  is of course obvious. Further experiments are required to ascertain the predicted broken symmetry \cite{Zaletel2020,Berg2021} in TBG.  TBWSe is however a textbook case for an AFM-XY model \cite{ChubukovV2025}.  The recent verification of the prediction of a specific heat $\propto T \ln T$ close to the fermion density where the loop-current order is extrapolated to turn on as $T \to 0$, gives direct evidence for a quantum-critical point in cuprates \cite{Tailleferspht, Klein2021}. The same singularity in the specific heat has earlier been observed at the antiferromagnetic criticality in heavy-fermion compounds \cite{HvLRMP2007} and in Fe-based compounds \cite{Hosono-rev}, which show a linear-in-$T$ resistivity and in the thermopower in all of them. An additional recent feature in all these compounds \cite{Analytis1, Shekhter2018} as well as in TBG \cite{Efetov2021} and TBWSe \cite{Pasupathy2021}  is that the resistivity is linear also in an applied magnetic field $|H|$ with magnitude such that the magnetoresistance at $\mu_B |H| = k_BT$ is similar to the zero-field resistance at $T$. The theory for this phenomenon and quantitative comparison with experiments has also been recently given \cite{CMV2022_R_H} based on the theory of the QXY-F model.

Calculations of transport properties at finite temperatures are a hard problem. Even in the problem of transport with electron-phonon interactions, in which everything essential was {\it understood} long ago by Peierls (for a historical review see \cite{Peierls1980}), an exact solution at low temperatures including Umklapp scattering has not been possible, although a detailed formalism was developed by Holstein \cite{Holstein1964}. For the Hubbard model in two dimensions, precise results including Umklapp scattering have been found only to second order in the interaction parameter~\cite{MFI, MFII}. 

We are able to present an exact low temperature theory for electrical and thermal conductivity as well as the Seebeck coefficient in the problem of transport with scattering of fermions by the fluctuations calculated for the QXY-F model but only for a circular Fermi surface, for which the single-particle self-energy is momentum independent for all momenta about the Fermi surface and where the energy varies linearly with momentum. One can then use relevant Ward-Takahashi identities to simplify calculations of the vertices. Even in that case, we can calculate  only because of the simplicity and unusual nature of the correlation function for this model \cite{Aji-V-qcf1, Aji-V-qcf3, Hou-CMV-RG}. The simplicity comes from the fact that the spectrum of fluctuations is a product of a function of frequency and of momentum. Moreover, the spatial correlation length normalized to the lattice constant is exponentially smaller than the temporal correlation length normalized to the short-time cutoff. These results are available from a precise quantum Monte Carlo calculation on a lattice including the renormalization of the critical fluctuations through coupling to fermions \cite{ZhuChenCMV2015, ZhuHouV2016} as well as a renormalization group (RG) calculation \cite{Hou-CMV-RG}. (A summary of the model and its solution are given in Appendix A.)

As argued by Peierls \cite{Peierls1980}, the electrical conductivity in a pure metal without Umklapp scattering is infinite at all temperatures. This is true both for a continuum model for a metal or a lattice model in which the current is proportional not to the momentum but to the group velocity of particles near the Fermi surface. For simple kinematic reasons Umklapp scattering is usually ineffective for fluctuations close to ${\bf q} = 0$ because of scarcity of low energy excitations of the order of  $k_BT$ at such ${\bf q}$. It leads to extra powers of temperature in transport relaxation rates compared to single-particle relaxation rates (for example $T^5$ in resistivity and $T^3$ in single-particle relaxation rates in electron-phonon scattering).  In the cuprates and most likely in the other quantum-critical problems with linear-in-$T$ resistivity, the single-particle relaxation rate has the same temperature dependence as the momentum relaxation rate which determines the resistivity.  It is then necessary for adequate phase space in scattering that there also be low energy excitations at large as well as small ${\bf q}$. This is one of the rather unique conditions met in the fluctuation spectrum derived for the QXY-F model summarized in Appendix A by Eq. (\ref{D}).  We show that Umklapp scattering in this case gives only a geometry dependent but temperature-independent numerical factor. Interestingly, we find that Umklapp scattering is not required for  relaxing energy current which determines the thermal conductivity. We give reasons based on symmetry why this is so. We derive that for a circular Fermi surface with wave number $k_F$ on a square lattice with a lattice constant $a_L$, $\frac{k_Fa_L}{\pi}$  must be larger than  about 0.552 for there to be finite resistivity in the pure limit.
  
 Else and Senthil \cite{ElseSenthil2021} have recently discussed the role of conservation laws on the conductivity at $T \to 0$ for a circular Fermi surface in the pure limit for which the only conservation is for momentum, density and energy. There are an infinite number of conservation laws at $T=0$ for a  general Fermi surface. In this paper, we  discuss conservation laws in the pure limit for a general Fermi surface and for $T \ne 0$. An infinite number of conservation laws exist even at $T \ne 0$ in the pure limit if there are regions on the Fermi surface in which Umklapp scattering is kinematically not allowed. In that case, the resistivity even at finite temperatures is zero. 

The paper is lengthy but the principal results are to be found in less than its first half. Figures~\ref{Fig:cond},~\ref{Fig:SE}, and~\ref{Fig:EFV} with detailed captions summarize diagrammatically the calculations done and Sec.~IV explain why it is possible, in leading order in temperature and frequency, to actually do such calculations.

The order of presentation in this paper is as follows: \\
(i) We begin with the well-known Kubo formula for transport properties in Sec. II. \\
(ii) The previously obtained results for the propagator of the fluctuations for the QXY-F model, obtained earlier by quantum Monte Carlo and diagrammatic techniques are summarized in Appendix A. We use these to calculate the self-energies of the fermions in Appendix B, where there are two important new results: (1) The vertex corrections to the self-energy are $0$ at the Fermi energy and the Fermi surface ${\bf p}_F$, with corrections of $O((p_0 - \mu)/E_F, {\bf p -p}_F)/p_F)$; (2) The self-energy is calculated with the derived momentum dependence of the fluctuations and of their coupling to fermions and shown to be unaffected to leading orders in frequency and momentum dependence. \\ (iii)
 The vertices to external fields in the Kubo equations are calculated in Sec. II.  This requires a new strategy because the propagator of QXY-F model, which is the input to such diagrammatic calculations, being topological, cannot be calculated by diagrammatic techniques. How the relevant quantities that are the input to such diagrammatic calculations are extracted is explained in Sec. II A.  In Secs. II A and II B, the Kubo equations are recast in terms of an equation for the velocity distribution functions $\Phi$ that makes it easy to introduce the memory matrix $M$ for calculations of transport. Since the memory matrix technique may be unfamiliar, its relation to the familiar many body vertices and self-energies is explicitly given in the text and the details of calculations using $M$ are relegated to an Appendix. The three distinct physical contributions to  $M$  calculated there and described in Sec. II B are because of self-energy of fermions, and two distinct types of vertices to the collective modes of the Maki-Thompson type and the Aslamazov-Larkin type. We discuss the conservation laws in Sec. II-C including those for small enough Fermi surfaces and their consequences. \\
(iv) In Sec. III, the transport coefficients are introduced. These parts are quite general. 
 We then use the properties of the QXY-F model in Sec.~IV to show that when the self-energy is independent of momentum, the vertices simplify so that one need only to calculate the vertex renormalization because of the Aslamazov-Larkin processes. They arise from considerations of momentum conservation
for electrical conductivity.  Some important cancellations using  Ward identities show that the logarithmically renormalized mass of MFL never appears in electrical conductivity or thermal conductivity.   As an example, the coefficient of the linear-in-$T$ resistivity and $T$-independent thermal conductivity are given in Sec. IV for circular Fermi surface of various sizes in a square lattice in terms of the coupling constant to the collective fluctuations.  We also show that unlike electrical conductivity and thermal conductivity, the Seebeck coefficient has the logarithmic-in-temperature mass enhancement. \\
(v)  We summarize the results in Sec. VI.   \\
(vi) 
In Appendices C--F, several technical details of the calculations whose results are presented in the main part of the paper are given.

\section{Conductivity} 

The Kubo formula~\cite{Kubo1957statistical} expresses the conductivity  $\sigma(\omega, T)$ in terms of the retarded current-current correlation function $\chi^R_{{\bf J, J}}(\omega, T)$ for zero-momentum transfer,  
 \be
 \label{Kubo0}
{\bf \sigma}(\omega, T) = e^2 ~\frac{\chi^R_{{\bf J, J}}(\omega, T)- {\chi_{{\bf J, J}}(0, T)}}{
i \omega}.
\ee
$\chi^R_{{\bf J, J}}(\omega, T)$ is the analytic continuation of
\be
\label{chi}
\chi_{{\bf J, J}}(i \omega_m) = - \frac{2}{\beta V}  \sum_n  
v_{{\bf p}x} G({\bf p}, i\epsilon_n + i \omega_m) 
 G({\bf p}, i\epsilon_n) \Lambda ({\bf p}, i\epsilon_n; i\omega_n),
\ee
as shown in Fig.~\ref{Fig:cond}. 
$e$, $\beta$, and $V$ are the elementary charge, the inverse temperature, and the volume of the system, respectively. $G$ is the renormalized single-particle Green's function, {$v_{{\bf p}x}$ is the $x$ component of the bare velocity, and $\Lambda$ is the renormalized vertex coupling to external electric field. 

The formula for conductivity can be written in the following transparent and familiar form (re-derived below),
\be
\label{Kubo-1}
\sigma(\omega,T) = \frac{2 e^2}{V} \sum_{\bf p} 
\left(- \frac{\partial f(\epsilon_{\bf p}^*)}{\partial \epsilon_{\bf p}^*}\right) v^*_{{\bf p}x} \Phi ({\bf p}, \omega).
\ee
Here $f(\varepsilon) = 1/(e^{\beta\varepsilon}+1)$, 
$v^*_{{\bf p}x}=\partial \epsilon_{\bf p}^*/\partial p_x$ is the velocity renormalized because of interactions, and $\Phi({\bf p}, \omega)$ is the velocity distribution function renormalized exactly for the effect of interactions. The latter is the appropriate Boltzmann distribution function.

A similar Kubo formula gives the results for the thermal conductivity $\kappa$. We will present results of evaluation of $\kappa$ as well as $\sigma$.

\begin{figure}[t]
\begin{center}
\includegraphics[width=0.5\columnwidth]{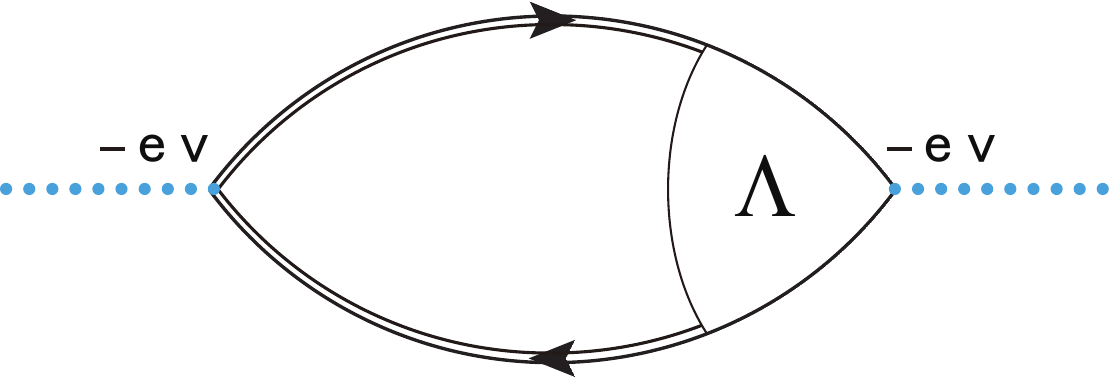}
\end{center}
\caption{Diagrammatic representation of the Kubo equation for the electrical conductivity. The external field coupling to fermion charge times their velocity is shown as a dotted line. The lines are exact single-particle Green's functions $G$. $\Lambda$ is the renormalized current vertex.}
\label{Fig:cond}
\end{figure}

The interesting and hard part of the calculation of the conductivity is the calculation of the external field (EF) - electron vertex $\Lambda$ or equivalently the velocity distribution function $\Phi$. We structure our calculation applying the Baym--Kadanoff conserving  scheme \cite{Holstein1964,Baym-K1961,Baym1962} to the QXY-F model and an extension to collective fluctuations of the Memory matrix formalism used in \cite{MFI, MFII}. The inter-relationship of the Memory matrix method to the more familiar many body techniques in quantum field theory~\cite{AGD, Nozieres-book} is given below. 

\subsection{Method for calculating self-energies and vertices} 

The quantum Monte Carlo calculations (QMC) and a renormalization group (RG) method [which is an extension of Kosterlitz's (RG) for the classical XY model] for the critical fluctuations of the quantum XY model on a lattice with coupling to fermions provides the collective fluctuation propagator $D$ including its self-energy through coupling to fermions and Umklapp scattering owing to the square lattice. 
The calculations of the fluctuations that are caused by topological excitations cannot be obtained by diagrammatic methods.  The method we use for evaluating the Kubo formula here is diagrammatic. So we need a strategy for how to extract the essential input for the calculations performed here from the QMC results. 
There are two essential inputs:\\

(a) {\it Irreducible vertex.} We need the irreducible vertex 
$I({\bf p},{\bf p}', i\epsilon_n, i\epsilon_{n'};i\omega_m)$ 
among the fermions through the interaction with the collective modes. Let $D({\bf q}, i\nu_l)$ be the propagator of the collective modes with which the fermions scatter with a coupling function  $g({\bf p,p +q})$. The bare irreducible vertex of the fermions caused by exchanging the collective mode fluctuations is 
given by 
\be
\label{bare-I}
{
I_0({\bf p},{\bf p}', i\epsilon_n, i\epsilon_{n'};i\omega_m) = |g({\bf p},{\bf p}')|^2 D({\bf p}-{\bf p}',i\epsilon_n-i\epsilon_{n'}).
}
\ee
The complete irreducible vertex is given in Fig.~\ref{Fig:SE}(a). 

\begin{figure}[t]
\begin{center}
\includegraphics[width=0.9\columnwidth]{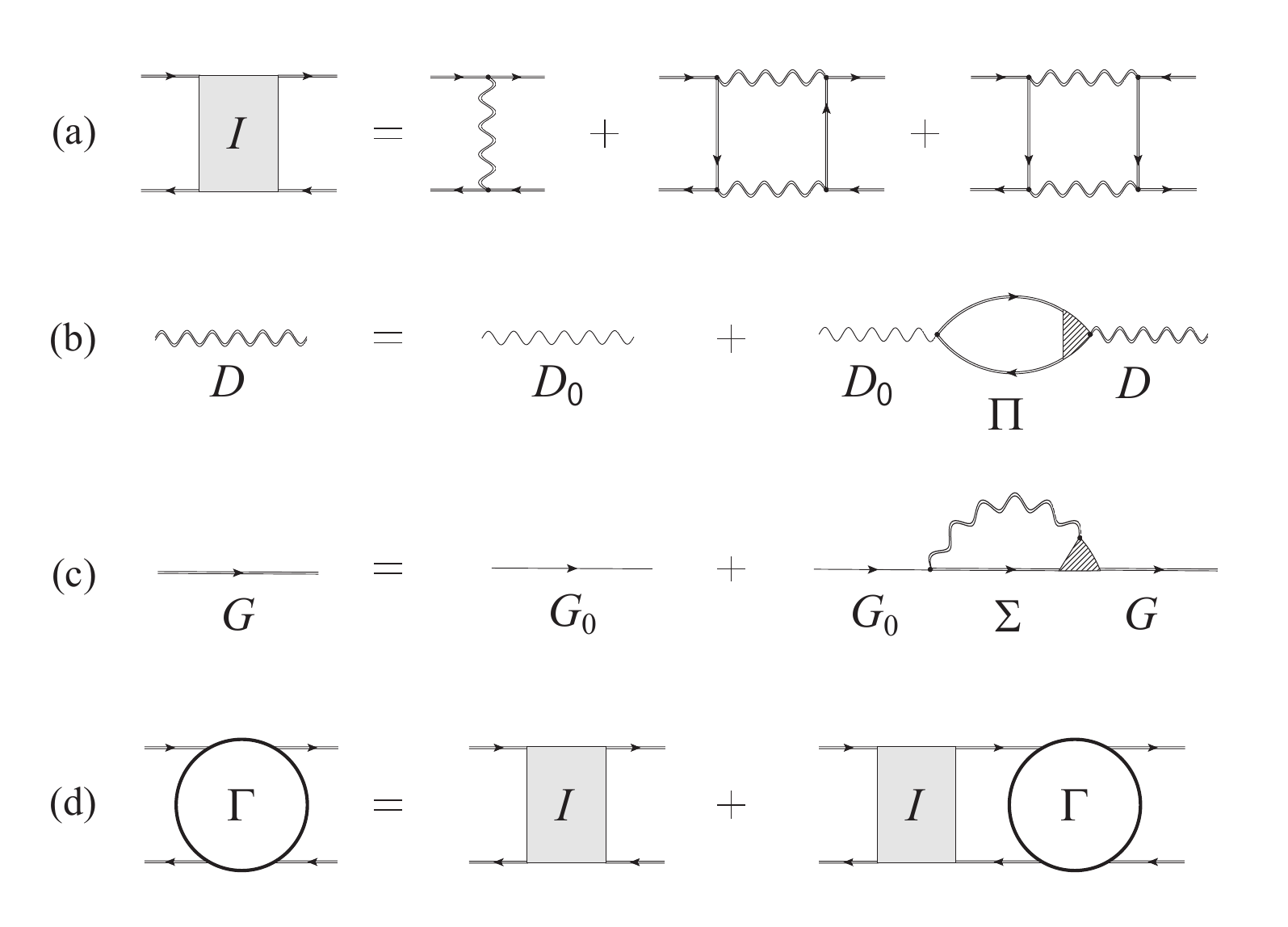}
\end{center}
\caption{Diagrammatic representation of (a) the irreducible vertex $I$ in terms of the bare-irreducible vertex $I_0$ which is given by the first diagram in which the wiggly line is given by (b), the renormalized collective mode propagator. In our case, the exact collective mode propagator [the left-hand side of Eq. (b)] is already available from quantum Monte Carlo calculation and renormalization group calculations for the quantum XY model coupled to fermions, summarized in Appendix A. $\Pi$ is the exact polarizability of the fermions which occurs in the self-energy of the collective mode propagator. $G$ is the exact fermion Green's function. It is calculable at momenta close to Fermi surface because the triangular vertex in its last part is $1$ to leading order in $\omega/E_F$ and $({\bf p-p}_F)/p_F$, as shown in Appendix B. (d) The total vertex that is used in the transport equations.}
\label{Fig:SE}
\end{figure}
 
The bare vertex in the microscopic theory of the Landau Fermi liquid is a constant. In contrast, the essential new physics for the class of non-Fermi liquids we are dealing with is the singularity in $D_R({\bf q},\omega)$ near a quantum-critical point. This singularity in momentum and frequency is exhibited in Eq. (\ref{Dqw}) of Appendix A.  $D_R({\bf q}, \omega)$ is of product form in frequency and momentum, unlike theories of the Ginzburg-Landau-Wilson type or their quantum generalizations.  At criticality, i.e. for correlation length $\xi_{\tau}$ in time and $\xi_r $ in space both $\to \infty$,
\be
\label{DR}
{\rm Im} D_R({\bf q},\omega) = - \chi_0\tanh(\omega/2T) \frac{1}{{\bf q}^2 a^2},
\ee
 with an upper cut-off in energy, which has been roughly estimated in Ref.~\cite{ASV2010} to be on the scale of 1/2 eV for the cuprates. $\chi_0$ is the amplitude of the fluctuations. The important changes away from criticality are given in Appendix; 
   the correlation length in space normalized to the lattice constant $\xi_r/a$ is related to the correlation length in imaginary time normalized to a short time cutoff $\xi_{\tau}/\tau_c$ by $(\xi_r/a)  = \log (\xi_{\tau}/\tau_c)$. The problem belongs to the class in which the dynamical critical exponent $z \to \infty$
\be
z = \frac{\partial \ln\xi_{\tau}}{\partial \ln\xi_r}.
\ee
We will show in Appendix B that the fermion self-energy depends smoothly on $\xi_r^{-1}$ even as it tends to $0$; so the problem remains effectively spatially local.

In Appendix A, we rederive  more physically the result \cite{ASV2010}  that in the continuum approximation for fermions, the relevant vertex 
coupling function  of the fermions to the critical modes,
\be
\label{g}
g({\bf p,p'}) \approx i \bar{g} ~ \big(\frac{{\bf p \times p}'}{m}\big) F(|{\bf p}|, |{\bf p}'|),
\ee
where $F(|{\bf p}|, |{\bf p}'|)$ is a dimensionless cutoff function of the order of the inverse lattice spacing and such that $\bar{g}$ is the dimensionless coupling constant.  Results are also available for fermions on a lattice \cite{ASV2010} and given in Appendix A. 

(b) {\it Fermion polarizability in the channel of the collective fluctuations.}
 Without the coupling to the fermions the collective mode propagator $D_0({\bf q}, \nu)$, which is given by Eq.~(\ref{D0}), 
is the propagator of the (2+1)D quantum XY model that is the same as the propagator of the 3D classical XY model.  
Therefore the nontrivial imaginary part in the full propagator $D_R({\bf q}, \nu)$ comes only from the coupling to fermions. 
This is shown diagrammatically in Fig.~\ref{Fig:SE}(b) and as the equation,
\be
\label{D}
D({\bf q}, i\nu_l) = D_0({\bf q}, i\nu_l) + D_0({\bf q}, i\nu_l) \Pi({\bf q}, i\nu_l)D({\bf q}, i\nu_l),
\ee
 or 
 \be
 \label{Dinv}
 D^{-1}({\bf q}, i\nu_l) = D_0^{-1}({\bf q}, i\nu_l) - \Pi({\bf q}, i\nu_l).
 \ee
 Since we know both $D_0$ and $D$ exactly, we know the polarizability $\Pi$ exactly. 
In particular, since $D_0$ has only an infinitesimal imaginary part, at criticality 
\be
{\rm Im} D_R^{-1}({\bf q}, \nu) = -{\rm Im} \Pi_R({\bf q}, \nu),
\ee
which may be re-written as 
 \be
\label{DRDA}
D_R (q) D_A (q) 
= \frac{{\rm Im} D_R (q)}{{\rm Im} \Pi_R (q)} .
\ee
Equation~(\ref{DRDA}) connects the fermion polarizability to the known collective mode propagator.
 It is important to note for calculating the vertex  that $D$ is a functional of the fermion propagator as may be seen from Fig.~\ref{Fig:SE}(b). The availability of the renormalized propagator of the fluctuations for the wiggly lines in Fig.~\ref{Fig:SE} so that they need not be calculated in the evaluation of the fermion vertices. 
 
\begin{figure}[t]
\begin{center}
\includegraphics[width=0.9\columnwidth]{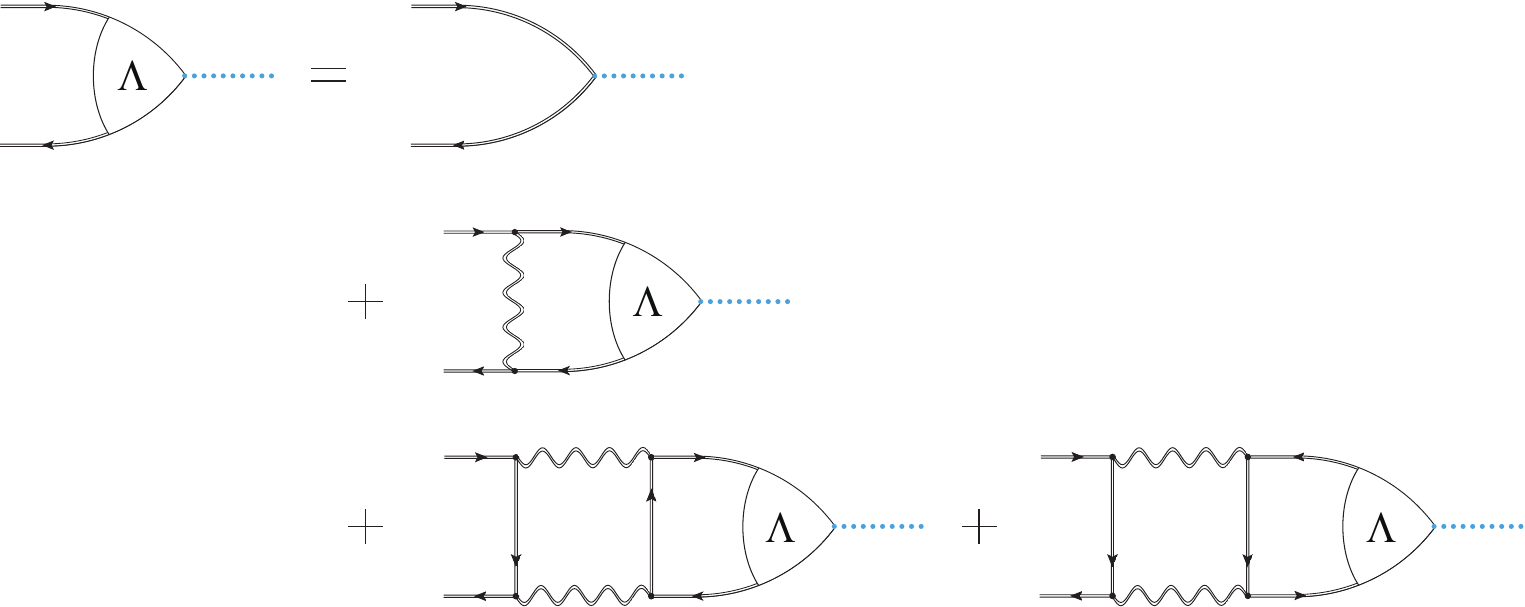}
\end{center}
\caption{Diagrammatic representation for the integral equation for the  external field vertices in calculation of conductivity. This is a sum of three parts shown successively in the three lines. First is the vertex coupling to the renormalized  fermion propagators, second and third are the vertex coupling to the collective modes. The second line gives what may be called "Maki-Thompson" diagrams and the third  corresponds to the "Aslamazov-Larkin" diagrams. We show that the first two give no correction for the present problem and the third comes from satisfying conservation laws and gives only a numerical correction }
\label{Fig:EFV}
\end{figure}
 
\subsection{Self-energies and vertices} 

The irreducible vertex $I$, represented in Fig.~\ref{Fig:SE}(a) (using which one can calculate the vertex $\Lambda$  as well as the fermion self-energy $\Sigma$) is given by the functional derivative \cite{Baym-K1961,Baym1962},
\be
\label{I}
I = \delta \Sigma/\delta G
\ee
Because the exact Green's function $D$ is a functional of $G$, the irreducible vertex $I$ includes contributions not only from what might be called the Maki-Tompson (MT) type  diagram (the second line in the diagrams shown in Fig.~\ref{Fig:EFV}) but also from the two Aslamazov-Larkin (AL) type diagrams (the third line in the digram), through the functional derivative of $D$ with respect to $G$.

In Appendix B, we show that for a circular Fermi surface, the self-energy is independent of momentum for deviation of momentum in which the energy varies linearly. This was true in the phenomenology but is also true based on the microscopic theory. Then it is possible to show that the MT type diagrams do not contribute for the present problem, but the AL diagrams are essential for providing the Umklapp factor. Also given the same condition, the first line of Fig.~\ref{Fig:EFV} gives no correction from the self-energy of the propagator and is therefore just the bare velocity. This as well as the absence of the MT diagram was noted earlier \cite{Shekhter-V-Hydro} but the contribution of the AL diagrams was not noted.}

\begin{figure}[t]
\begin{center}
\includegraphics[width=0.95\columnwidth]{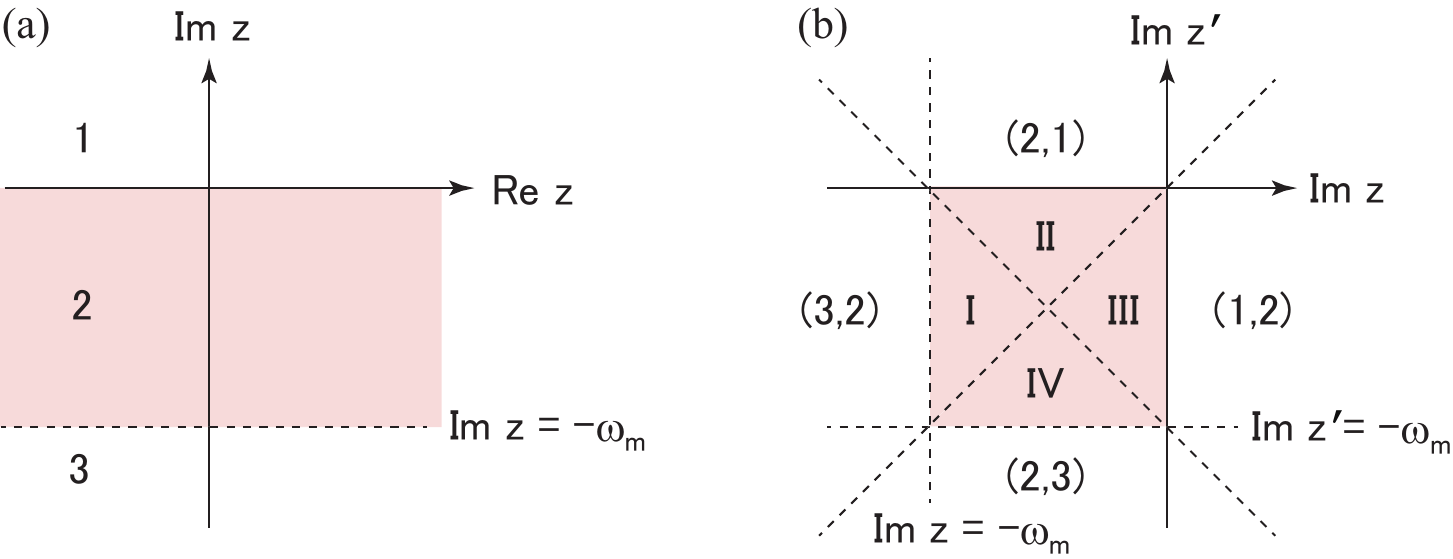}
\end{center}
\caption{Regions on the complex energy plane for the analytic continuations of (a) $\Lambda({\bf p}, z; i\omega_m)$ with $i \epsilon_n = z$ and (b) $I({\bf p}, i \epsilon_n, {\bf p}', i \epsilon_{n'}; i \omega_m)$ with $i \epsilon_n = z$ 
and $i \epsilon_{n'} = z'$.}
\label{Fig:analytic}
\end{figure}

We will need the analytic continuation of the EF electron vertex part given by an integral equation represented in Fig.~\ref{Fig:EFV}. This is accomplished following the classic paper by \`{E}liashberg \cite{Eliashberg1961}.  
By taking $i\omega_m \to \omega + i0^+$ with $\omega = 0$ after the analytic continuation of $\Lambda({\bf p}, i\epsilon_n; i\omega_m)$ to the region 2 shown by Fig.~\ref{Fig:analytic}(a) on the complex energy plane of $i \epsilon_n = z$, we obtain 
\be
\Lambda_2 (p) &=& 
{\tilde v}_x (p)  
+ \frac{1}{2\pi}\cosh \frac{\varepsilon}{2T}  
\int_{p'}\, 
{\cal I}(p,p') \,{\rm sech} \frac{\varepsilon'}{2T} \quad
\nonumber \\
&\quad& \times G_R (p') G_A (p')
\Lambda_2 (p'),
\label{Lanbda2}
\ee 
where $\int_{p'} = V^{-1}\sum_{{\bf p}'}\int_{-\infty}^{\infty} d\varepsilon'$.
The first term on the right comes from the diagrams including $G_RG_R$ and $G_AG_A$ and they satisfy the Ward identity or the continuity equation, 
\be
{\tilde v}_x (p) = v_{{\bf p}x} + \frac{\partial\,{\rm Re} \Sigma_R (p)}{\partial p_x} .
\ee
The second term comes from  the particle-hole sections with  $G_RG_A$ and 
${\cal I}(p,p')$ is given by
\be
{\cal I}(p,p') &=& - \frac{1}{2i} \left[ I_{22}^{\,{\rm I\hspace{-0.1em}I}} (p,p') - I_{22}^{\,{\rm I\hspace{-.12em}I\hspace{-.12em}I}} (p,p') \right] 
{\rm cosech} \frac{\varepsilon - \varepsilon'}{2T}
\nonumber \\
& & + \frac{1}{2i} \left[I_{22}^{\,{\rm I\hspace{-.12em}I\hspace{-.12em}I}}(p,p') 
- I_{22}^{\,{\rm I\hspace{-0.12em}V}}(p,p') \right] 
{\rm cosech} \frac{\varepsilon + \varepsilon'}{2T} , \quad
\label{calI_analytic}
\ee
where $I_{22}^{\,{\rm I\hspace{-0.1em}I}}$, $ I_{22}^{\,{\rm I\hspace{-.12em}I\hspace{-.12em}I}}$, and $I_{22}^{\,{\rm I\hspace{-0.12em}V}}$ are the analytic continuations of the irreducible vertex 
$I({\bf p}, i \epsilon_n, {\bf p}', i \epsilon_{n'}; i \omega_m)$ to the regions shown by Fig.~\ref{Fig:analytic}(b) with $i \epsilon_n = z$ and $i \epsilon_{n'} = z'$. 

Adding the contributions from the three processes shown in Fig.~\ref{Fig:SE}(a), Eq.~(\ref{calI_analytic}) leads to
\be
{\cal I}(p,p') &=& - |g({\bf p}, {\bf p}')|^2 {\rm Im} D_R (p\!-\!p') {\rm cosech} \frac{\varepsilon \!-\! \varepsilon'}{2T}
\nonumber \\
&& + \pi \int_{p_1}\,
|g({\bf p},{\bf p}_1)|^4 D_R (p\!-\!p_1) D_A (p\!-\!p_1) 
\nonumber \\
& & \times \left(
A (p_1) A (p_1\!-\!p\!+\!p') 
 {\rm sech} \frac{\varepsilon_1}{2T}   {\rm sech} \frac{\varepsilon_1\!-\!\varepsilon\!+\!\varepsilon'}{2T}  \right.
\nonumber \\
& & - \left.
A (p_1) A (p\!+\!p'\!-\!p_1) 
 {\rm sech} \frac{\varepsilon_1}{2T}   {\rm sech} \frac{\varepsilon\!+\!\varepsilon'\!-\!\varepsilon_1}{2T}  \right). 
 \label{calI1}
\ee
$A(p)$ is the single-particle spectral function,
\be
A(p) = - \frac{1}{\pi} {\rm Im} G_R (p). 
\ee
The conservation of charge leads to a generalized Ward--Takahashi identity, which defines the irreducible vertex Eq. (\ref{I}) and relates the imaginary part of the fermion self-energy to the irreducible vertex as
\be
2 {\rm Im} \Sigma_R (p) \, {\rm sech} \frac{\varepsilon}{2T} 
&=& - \int_{p'}\, {\cal I}(p,p') A (p')  {\rm sech} \frac{\varepsilon'}{2T} .
\label{Ward--Takahashi}
\ee
The energy and temperature dependences of the imaginary part of the self-energy is derived in 
Appendix B and is given to leading order with corrections of 
$O(\varepsilon/E_F, |{\bf p}-{\bf p}_F|/p_F)$ by
\begin{align}
{\rm Im} \Sigma_R({\bf p},\varepsilon) 
&= - \bar{g}^2 N(\varepsilon)\, \varepsilon \coth \frac{\varepsilon}{2T} ,
\label{result_self-energy} 
\end{align}
where $N(\varepsilon) = V^{-1}\sum_{\bf p} A({\bf p},\varepsilon)$
is the density of states of fermions per spin. 
As we will see in more detail below, the momentum independence of the self-energy leads to the MT contribution to $I$ to be zero. 
For self-energy independent of momentum, the AL contribution is finite 
but suppressed by Umklapp scattering, in which the initial momentum and the final momentum differ by reciprocal vectors. We are interested only in the self-energy in the first Brillouin zone. So only the first contribution to the irreducible vertex in Fig.~\ref{Fig:SE}(a) 
and Eq.~(\ref{calI1}) contributes to the integral for the self-energy given by (\ref{Ward--Takahashi}). Then the integral in this equation can be evaluated. 

By use of ${\tilde v}_x (p)$ and $\Lambda_2 (p)$, the Kubo formula, Eq. (\ref{Kubo0}), can be written for $\omega=0$ as
\be
\sigma(T) = \frac{e^2}{\pi} \int_{p}\,
\left( - \frac{\partial f(\varepsilon)}{\partial \varepsilon} \right) 
{\tilde v}_x (p) G_R (p) G_A (p)\Lambda_2 (p) ,
\label{Kubo-Lambda2}
\ee
where $-\partial f(\varepsilon)/\partial \varepsilon = (1/4T)\,{\rm sech}^2 (\varepsilon/2T)$. 
Noting that 
\be
G_R (p) G_A (p) 
= \frac{{\rm Im} G_R (p)}{{\rm Im} \Sigma_R (p)}, 
\label{Phi}
\ee
we can rewrite Eq. (\ref{Kubo-Lambda2}) so that electrical conductivity, including its $\omega$ dependence is given by Eq. (\ref{Kubo-1})
as promised earlier,
\be
\sigma(\omega, T) = 2 e^2 \int_{p}\,
\frac{1}{4T} \,{\rm sech}^2 \frac{\varepsilon}{2 T}
{\tilde v}_x (p) A (p) \Phi (p,\omega) .
\label{Kubo-Phi}
\ee
The electron velocity distribution function $\Phi(p)$ is 
\be
\Phi (p)  = - \frac{\Lambda_2 (p)}{2 {\rm Im} \Sigma_R (p)}.
\label{n}
\ee

\subsection{Conservation laws and transport using memory matrix} 

Everything so far has been simple, but nothing has been evaluated. To do so, it is convenient to introduce the "memory matrix" $M(p,p')$ 
by  rewriting Eq.~(\ref{Lanbda2}) in terms of  $\Phi(p)$ instead of $\Lambda_2(p)$ 
\be
\int_{p'}\, 
M''(p,p')  A(p') \Phi(p') \,{\rm sech} \frac{\varepsilon'}{2T} 
= {\tilde v}_x (p)  \,{\rm sech} \frac{\varepsilon}{2T}.
\label{Lambda_to_Phi}
\ee 
By comparing Eqs.~(\ref{Lanbda2}) and (\ref{Lambda_to_Phi}), we obtain the imaginary part of the $M$,
\be
M''(p,p')  = 
\frac{\Delta\Sigma (p)}{\Delta G(p')}\delta(p-p') -  {\cal I}(p,p') 
\label{}
\ee 
with $\delta(p-p') = V \delta_{{\bf p},{\bf p}'} \times 2 \pi \delta(\varepsilon-\varepsilon')$, 
$\Delta \Sigma(p) = \Sigma_R(p)-\Sigma_A(p)$, and $\Delta G(p) = G_R(p)-G_A(p)$. 

Using Eq.~(\ref{Lambda_to_Phi}), one can derive the equation for the static velocity distribution function $\Phi({\bf p},\omega =0)$. 
This is sufficient to derive the dc transport coefficients. To discuss the frequency dependent conductivity and the conservation laws, we use the result in Eq. (2.35) of Ref. \cite{MFI} to linear order in $\omega$. [The  two terms explicitly  linear in $\omega$ on the right side of Eq. (\ref{Lambda3}) below are the modifications beyond what is needed for the static transport properties]. We will not reproduce the derivation given  in \cite{MFI}  but simply state the result, 
\be
{\tilde v}_x (p) \, {\rm sech} \frac{\varepsilon}{2T}&=& \left[-i\omega a^{-1}(p) + 2 {\tilde \Gamma}(p) \right]  \Phi(p, \omega)\, {\rm sech} \frac{\varepsilon}{2T}
  \nonumber
\\
&\quad&-\int_{p'}\, \left[-i\omega \Gamma^k(p,p') \frac{1}{4T} 
  \, {\rm sech} \frac{\varepsilon}{2T}\,{\rm sech}\frac{\varepsilon'}{2 T}
+  {\cal I}(p,p')  \right]
A(p') \Phi (p',\omega) \, {\rm sech} \frac{\varepsilon'}{2T}, 
\nonumber\\
&&
\label{Lambda3}
\ee
where $a(p)^{-1}=1-\partial  {\rm Re} \Sigma_R(p)/\partial \varepsilon$, 
${\tilde \Gamma}(p)= - {\rm Im} \Sigma_R(p)$, and $\Gamma^k(p,p')$ is the full four-point vertex
in the $k \to 0$ after $\omega \to 0$ limit. $\Gamma^k(p,p')$ is the familar full vertex that is written 
[see Fig.~\ref{Fig:SE}(d)], in terms of the complete irreducible vertex 
and the single particle Green's function. 
As usual, it can also be rewritten in terms of 
$\Pi_R ({\bf q}, \nu)$, 
given diagrammatically in Fig.~\ref{Fig:SE}(b)~\cite{AGD}.

Equation~(\ref{Lambda3}) is exact and in terms of quantities familiar in traditional many body theory \cite{AGD}. It can be reexpressed in a simpler looking form, as shown in Ref.~\cite{MFI},
in terms of the memory matrix,  as a generalizaton of Eq. (\ref{Lambda_to_Phi}) including the frequency dependence,
\be
{\tilde v}_x (p)  \,{\rm sech} \frac{\varepsilon}{2T} = -i \int_{p'}\, 
M(p,p';\omega)  A(p') \Phi(p') \,{\rm sech} \frac{\varepsilon'}{2T}.
\label{}
\ee

To make the memory matrix familiar, we reproduce the derivation of the well known Landau-Boltzmann transport equation based on it in Appendix F. 
Transport coefficients of interest are written in terms of it in Sec. \ref{S-trans} in their full generality and used to evaluate them for the QXY-F model in Sec. \ref{S-mfl}.

The memory matrix method is most convenient to discuss the conservation laws for transport. An essential summary of the memory matrix technique for transport at arbitrary frequency and temperature is given in Refs. \cite{MFI, MFII}. We define it here in terms of the single-particle spectral function $A$, 
the complete vertex $\Gamma$ and the irreducible vertex $I$, familiar in traditional many-body theory and already introduced and exhibited in Fig.~\ref{Fig:SE}.

Using the dimensionless energy variables $t$ scaled by temperature, i.e. $\varepsilon \to T t$ and omitting the momentum variables for brevity, we use an abbreviated notation in which 
\be
A(t) \equiv A({\bf p},Tt); ~~
{\tilde \Gamma}(t) \equiv {\tilde \Gamma}({\bf p},Tt), ~~ {\cal I}(t,t') = {\cal I}({\bf p},Tt,{\bf p}',Tt'). 
\ee
We can then write Eq.~(\ref{Kubo-Phi}) and (\ref{Lambda3}) as
\be
\sigma(\omega,T) &=& 2 \int_{-\infty}^{\infty}  dtX_1(t) A(t) \Phi_1(t).
\label{L11omega}
\ee
where 
\be
\Phi_1(t) &=& -e \Phi({\bf p}, Tt)\frac{1}{2}\,{\rm sech} (t/2),\\
X_1(t) &=& -e {\tilde v}_x({\bf p}, Tt)\frac{1}{2}\,{\rm sech} (t/2) \\
&=& -i\omega \Phi_1(t) - i \int_{-\infty}^{\infty} dt' M(t,t') A(t') \Phi_1(t') ,\qquad
\label{BS-memory}
\ee
By comparison with Fig. \ref{Fig:SE} for conductivity and the equations immediately following, one can deduce that
\be
\label{ImMsimplified}
M'(t,t') &=& \omega \left( a^{-1} - 1 \right) A^{-1}(t)  \delta(t-t')
- \omega \Gamma^k \frac{1}{4} \,{\rm sech} \frac{t}{2} \,{\rm sech} \frac{t'}{2} ,
\\
\label{memory1}
M''(t,t') &=& 2{\tilde \Gamma}(t) A^{-1}(t) \delta(t-t') - T {\cal I}(t,t').
\ee 
In Eqs.~(\ref{L11omega}) and (\ref{BS-memory}), the sum with respect to the momentum is implicit. The equivalence of Eq. (\ref{L11omega}) with Eq. (\ref{Kubo-Phi}) can be easily verified through direct substitution.

From Eq.~(\ref{BS-memory}), 
\be
\Phi_1(t) =i \int dt' [\omega \delta (t-t') + M(t,t') A(t')]^{-1} X_1(t'), 
\ee
so that the electrical conductivity can be written using the memory matrix,
\be
\sigma(\omega,T) &=& 2 i \int_{-\infty}^{\infty} dt d t' X_1(t) 
[\omega A^{-1}(t) \delta (t-t') + M(t,t')]^{-1} X_1(t') .\qquad
\label{sigma_genT}
\ee
Now we note that for low energies and temperatures, for any metal obeying Luttinger's theorem, whether Fermi liquid or marginal Fermi liquid
\be
A({\bf p}, \epsilon) = a(\epsilon) \delta\big(\epsilon - \epsilon({\bf p})\big) = \delta \big(a^{-1}(\epsilon) \epsilon - \tilde{\epsilon}({\bf p})\big) \equiv \delta\big(a^{-1} T t -  \tilde{\epsilon}({\bf p})\big).
\ee
For a Fermi liquid $a^{-1}$ is a constant and for a marginal Fermi liquid $a^{-1} \propto \ln T$. Therefore at low temperatures, $A(t) = A({\bf p}, Tt)$ in Eqs.~(\ref{memory1})--(\ref{sigma_genT}) can be replaced by $A_0 \equiv A({\bf p}, 0) = \delta ({\tilde{\epsilon}}_{\bf p})$, 
where ${\tilde{\epsilon}}_{\bf p} = {\epsilon}_{\bf p} + {\rm Re} \Sigma_R  ({\bf p}, 0)$ with ${\epsilon}_{\bf p}$ giving the noninteracting energy dispersion relative to the chemical potential. All the functions with this replacement are then distinguished by the subscript $0$. 

Thus, at the lowest order in temperature, the momentum summation is restricted even for a marginal Fermi liquid (or a more singular Fermi liquid) on the Fermi surface. Now we define the inner product in this restricted momentum space as 
\be
\langle u | v \rangle &=& (2\pi)^{-d} \int d{\bf p}\, u_{\bf p} v_{\bf p} \delta ({\tilde{\epsilon}}_{\bf p})/N(0); 
N(0) = (2\pi)^{-d} \int d{\bf p} \,\delta ({\tilde{\epsilon}}_{\bf p}), \\
{\hat M} |u\rangle &=& (2\pi)^{-d} \int d{{\bf p}'}M_{{\bf p}{\bf p}'}u_{{\bf p}'} \delta ({\tilde{\epsilon}}_{{\bf p}'})/N(0). 
\ee
Then the leading term of Eq.~(\ref{sigma_genT}) with respect to temperature is given by
\be
\sigma(\omega,T) &=& 2 i e^2 N(0) \int_{-\infty}^{\infty} dt d t' 
\langle {\tilde v}_x |[\omega \hat{1} \delta (t-t') + {\hat M}_0(t,t')]^{-1}|{\tilde v}_x \rangle
\frac{1}{4}\, {\rm sech} \frac{t}{2} \, {\rm sech} \frac{t'}{2},
\label{optical conductivity}
\ee
where $|{\tilde v}_x\rangle$ represents the vector for ${\tilde v}_{{\bf p}x} = \partial {\tilde{\epsilon}}_{\bf p}/\partial p_x$ in the defined inner product space. 

Equation~(\ref{optical conductivity}) manifests the conservation laws in the pure limit for $\sigma(\omega, T)$ quite generally. As pointed out in \cite{MFI}, the imaginary part of the memory matrix has zero eigenvalues when the system described by a Hamiltonian $H$ has conserved quantities. 
This is because a conservation law $[Q, H] = 0$ with $Q= \sum_{{\bf p}, \sigma}\theta_{\bf p} c^\dagger_{{\bf p}, \sigma} c_{{\bf p}, \sigma}$ leads to the generalized Ward-Takahashi identity,
\be
\label{Ward--Takahashi2}
2 {\rm Im} \Sigma_R (p) \theta_{\bf p}\, {\rm sech} \frac{\varepsilon}{2T} 
&=& - \int_{p'}\, {\cal I}(p,p') A (p') \theta_{{\bf p}'} {\rm sech} \frac{\varepsilon'}{2T} . 
\ee
In a lattice, the total momentum is not a conserved quantity if Umklapp scattering is allowed. At low temperatures, however, all the momentum involved in scattering must be on the Fermi surface, so Umklapp scattering is ineffective in a system with a small enough Fermi surface. In that case, $\theta_{\bf p} = p_x$ and we have
\be
\int_{-\infty}^{\infty} d t' \hat{M}''_0(t, t')|\theta \rangle \, {\rm sech} \frac{t'}{2} = 0.
\label{zeroeigenvalue}
\ee
Since $p_x$ has an overlap with ${\tilde v}_{{\bf p}x}$ on the Fermi surface, the resistivity vanishes and Eq.~(\ref{optical conductivity}) takes the form of $\sigma(\omega,T) = D^T /(- i \pi \omega)$ at low but nonzero temperatures. Note that $D^T$ differs from the usual Drude weight $D^\omega$ that is defined at zero temperature~\cite{MFI}. From Eq.~(\ref{memory1}), $D^T= 4\pi e^2 \langle {\tilde v}_x |\theta \rangle^2 /\chi_{\theta\theta}$ with $\theta_{\bf p} = p_x$, where $\langle {\tilde v}_x |\theta \rangle = n/2$ with the electron number density $n$ and $\chi_{\theta\theta}$ is the static limit of the momentum autocorrelation function, as pointed out in \cite{ElseSenthil2021}.

\subsection{Application to critical $\omega/T$ fluctuations} 

As will be shown in Sec. IV, if the critical fluctuations scale as a function of $\omega/T$, in other words, when the imaginary part of the propagator of the collective modes is as  ${\rm Im} D_R({\bf q}, \omega) = {\cal F}_{\bf q} (\omega/T)$, the imaginary part of the memory matrix is proportional to temperature, i.e., $\hat{M}''_0(t, t') \propto T$. If there is a conserved quantity so that $\hat{M}_0''(t, t')$ has a zero eigenvalue as described by Eq.~(\ref{zeroeigenvalue}), $\sigma(\omega,T) \propto  -i/\omega$ with a vanishing linear-in-$T$ resistivity. This is true for small Fermi surfaces where Umklapp is ineffective. However, as will be demonstrated in Sec. IV C, for large Fermi surfaces where Umklapp scattering occurs, there is no such conserved quantity and a nonzero linear-in-$T$ resistivity results. Since the real part $\hat{M}'_0(t, t')$ of the memory matrix is proportional to $\omega$, Eq.~(\ref{optical conductivity}) shows that the electrical conductivity then scales as a function of $\omega/T$ and may be written as $\sigma(\omega, T) = T^{-1} {\cal G}(\omega/T)$ with ${\cal G}(0)$ finite. 
Therefore, when the Fermi surface is large and the critical fluctuations obey a $\omega/T$ scaling, the quantum-critical conductivity also obeys a $\omega/T$ scaling.

\section{Low-temperature electrical and thermal conductivities} 
\label{S-trans}

The dc-electrical conductivity $\sigma$ and the thermal conductivity $\kappa$ can be written using the transport coefficients $L_{ij}$ as
\be
\sigma &=& L_{11}, \\
\kappa &=& \frac{1}{T} \left( L_{22} - \frac{L_{12}L_{21}}{L_{11}} \right).
\ee
Here $L_{11}$ is given by Eq.~(\ref{L11omega}) for $\omega=0$.
$L_{12} = L_{21}$ and $L_{22}$ are usually derived  from the Boltzmann transport theory~\cite{ziman2001electrons}, 
but for strongly correlated electron systems with short-range Coulomb interactions they are justified on the basis of the Kubo-Luttinger formalism~\cite{luttinger1964theory,PhysRevB.67.014408,ogata2019range}.
Introducing 
\be
X_2(t) = T {\tilde v}_x({\bf p}, Tt)(t/2)\, {\rm sech} (t/2)
\ee
 in addition to $X_1(t)$ defined above, 
these transport coefficients can be written together as
\be
L_{ij} &=& 2 \int_{-\infty}^{\infty}  dtX_i(t) A(t) \Phi_j(t),
\ee 
where $\Phi_i(t)$ ($i=1,2$) is the solution of the integral equation, 
\be
X_i(t) &=& \int_{-\infty}^{\infty} dt' M''(t,t') A(t') \Phi_i(t') 
\nonumber \\
&=& 2 {\tilde \Gamma}(t) \Phi_i(t)  - T \int_{-\infty}^{\infty} dt' {\cal I}(t,t') A(t') \Phi_i(t') .\qquad
\label{originalLBtransport}
\ee
This equation for $i=1$ corresponds to Eq.~(\ref{BS-memory}) with $\omega = 0$.

Now we assume that the low-temperature expansion of the spectral function $A(t)$ can be written as $A(t) = A_0 + A_1(t) + \cdots$ with $A_0 = \delta({\tilde \epsilon}_{\bf p})$. 
Then, applying this expansion to the spectral functions and to (\ref{calI1}), the self-energy and vertex parts can be expanded as 
${\tilde \Gamma}(t) = {\tilde \Gamma}_0(t) + {\tilde \Gamma}_1(t) + \cdots$ 
and $ {\cal I}(t,t')  =  {\cal I}_0(t,t')  +  {\cal I}_1(t,t')  + \cdots$, respectively. 
Let us write $Y_i(t)$ for the leading term in the low-temperature expansion of $\Phi_i(t)$, 
which satisfies 
\be
X_i(t) = \int_{-\infty}^{\infty} dt' M_0''(t,t') A_0 Y_i(t') 
= 2 {\tilde \Gamma}_0(t) Y_i(t) - T \int_{-\infty}^{\infty} dt' {\cal I}_0(t,t')A_0 Y_i(t'). \quad
\label{Ydiag}
\ee
Then the leading term of the diagonal transport coefficient is simply given by 
\be
L_{ii} =2 \int_{-\infty}^{\infty} dt  X_i(t) A_0 Y_i(t).
\label{Ldiag}
\ee
However, $X_1(t)$ and $Y_2(t)$ are even and odd functions of $t$, respectively, so $\int_{-\infty}^{\infty} dt  X_1(t) A_0 Y_2(t)$ $=$ $0$.
Hence the off-diagonal transport coefficient requires a higher-order expansion, and its leading term takes a rather complicated form,
\be
L_{12} &=&
4 \int_{-\infty}^{\infty} dt Y_1(t) {\tilde \Gamma}_0(t) A_1(t) Y_2 (t)
- 4\int_{-\infty}^{\infty} dt Y_1(t) {\tilde \Gamma}_1(t) A_0 Y_2 (t)
\nonumber \\
&\quad& + 2 T \int_{-\infty}^{\infty} dtdt' Y_1(t) A_0{\cal I}_1(t,t')A_0 Y_2 (t').
\label{L12general}
\ee
Since $L_{12}$ = $L_{21}$ arises from the higher-order terms in the low-temperature expansion, $L_{12}L_{21} \ll L_{11}L_{22}$, and thus the thermal conductivity is given by $\kappa = L_{22}/T$ at low temperatures.
Then, from Eqs.~(\ref{Ydiag}) and (\ref{Ldiag}), $\kappa$ as well as $\sigma$ can be obtained using the imaginary part  $M_0''(t,t')$ of the memory matrix whose momenta are limited on the Fermi surface because of the presence of $A_0 = \delta({\tilde \epsilon}_{\bf p})$.

For specific calculations of $\sigma$ and $\kappa$, it is useful to introduce the Fermi surface harmonics $\psi_L({\bf p})$ which are orthonormalized as
$\langle \psi_L | \psi_{L'}\rangle = (2\pi)^{-d}\int d {\bf p} \psi_L({\bf p}) \psi_{L'} ({\bf p})\delta({\tilde{\epsilon}}_{\bf p})/N(0) = \delta_{L,L'}$ and complete $\sum_L |\psi_L\rangle \langle \psi_L| = \hat{1}$. 
Since the bare vertex is an odd function of momentum, only the harmonics satisfying $\psi_L(-{\bf p}) = - \psi_L({\bf p})$ can form a basis and $\psi_L({\bf p})$ with $L=1$ is chosen to be proportional to ${\tilde v}_{{\bf p}x}$, i.e., $\psi_1({\bf p}) = {\tilde v}_{{\bf p}x}/\langle {\tilde v}_x^2 \rangle^{1/2}$ where $\langle {\tilde v}_x^2 \rangle = \langle {\tilde v}_x| {\tilde v}_x \rangle$ is the average of ${\tilde v}_x^2 $ on the Fermi surface. 
Using this basis of the Fermi surface harmonics and denoting the elements of the memory matrix as $[\hat{M}(t, t')]_{LL'} \equiv \langle \psi_L| \hat{M}(t, t')|\psi_{L'}\rangle$, Eq.~(\ref{optical conductivity}) with $\omega = 0$ can be written as
\be
\sigma &=& 
2 e^2N(0) \langle {\tilde v}_x^2 \rangle 
\int_{-\infty}^{\infty} d t d t' 
\frac{[ \hat{M}_0''(t, t')^{-1} ]_{11}}{4 \cosh(t/2) \cosh(t'/2)} . \quad\,\,
\label{sigma_FSH}
\ee 
From the consideration in the previous paragraph, the thermal conductivity at low temperatures can also be written in a similar form,
\begin{align}
\kappa &= 
2 T N(0) \langle {\tilde v}_x^2 \rangle
\int_{-\infty}^{\infty} dt dt'
\frac{[\hat{M}_0''(t,t')^{-1}]_{11} t t' }{4 \cosh (t/2) \cosh (t'/2)}.
\label{kappa_memory}
\end{align} 
From Eq.~(\ref{memory1}) with Eqs.~(\ref{calI1}) and (\ref{Ward--Takahashi}), the imaginary part of the memory matrix can be obtained as
\be
\hat{M}''_0(t, t') = \hat{M}''_{\rm SE}(t, t')+\hat{M}''_{\rm MT}(t, t')+\hat{M}''_{\rm AL}(t, t'),\quad
\ee
where the first,  second, and third terms are the contributions from the self-energy, MT-vertex, and AL-vertex corrections, respectively, (from which the subscripts $0$'s are omitted for brevity).

The matrix elements of the three contributions are given in Appendix C. From the expressions given, we note
 that the AL contribution is generally particle-hole asymmetric, which is essential for the Seebeck coefficient discussed in the next section, but its leading term given by Eq.~(\ref{memory_AL_FSH}) in the low-temperature expansion has particle-hole symmetry expressed as $\hat{M}''_{\rm AL}(t,t') = \hat{M}''_{\rm AL}(t,-t')= \hat{M}''_{\rm AL}(-t,t')$, while the bare vertex for $\kappa$ is an odd function of the energy variable. Thus the AL vertex corrections do not contribute to the leading term in $\kappa$, but they do for $\sigma$.

\section{Transport  for criticality of the quantum XY-model coupled to fermions} 
\label{S-mfl}

We  now use the developments in Sec. III  for the critical fluctuations and their coupling to fermions summarized in Appendix A, which are valid for fluctuations of any vector field obeying $U(1)$ symmetry in the quantum-critical region and coupled to fermions as in the model solved in \cite{Aji-V-qcf1, ZhuChenCMV2015, ZhuHouV2016, Hou-CMV-RG}. Let us first note from the expressions for the matrix elements, that when the imaginary part of the propagator of the collective modes is written as  ${\rm Im} D_R({\bf q}, \omega) = {\cal F}_{\bf q} (\omega/T)$, the imaginary part of the memory matrix is proportional to temperature, i.e., $\hat{M}''_0(t, t') \propto T$. Note that the eigenfunction with the zero eigenvalue described by Eq.~(\ref{zeroeigenvalue}) is an even function of the energy variable $t'$, which has an effect only on $\sigma$. On the other hand, the eigenvalues of $\hat{M}''_0(t, t')$ for eigenfunctions that are odd functions of the energy variable are nonzero positive in general. This is related to the symmetry that the AL vertex corrections are absent for $\kappa$. As a result, $\kappa$ is finite even without Umklapp scattering and is a temperature-independent constant from Eq.~(\ref{kappa_memory}) 
when $\hat{M}''_0(t, t') \propto T$.

There are two attributes that make the calculations easy. First is the $\omega/T$ dependence of the fluctuation spectra which determines the irreducible interaction $I_0$. 
The second is that the self-energy of the fermions is then effectively momentum independent, 
as shown in Eq.~(\ref{result_self-energy}). 
Therefore, by Eq.~(\ref{calI1}) and (\ref{Ward--Takahashi}), the irreducible interaction 
can also be treated as effectively momentum independent,
\begin{align}
|g({\bf p,p+q})|^2{\rm Im} D_R  ({\bf q}, \omega) 
= - \bar{g}^2 \tanh \frac{\omega}{2T}.
\label{assumption_fluctuation}
\end{align}

The essential technical aspects to calculate the transport coefficients has been reduced to evaluation of the Memory matrix $M(t,t')$.
We present their algebraic evaluation in Appendix C. Here we give the physical basis for the answers and collect the results for the transport coefficients derived there.

Noting from Eq. (\ref{BS-memory}) that $\Phi_1(p)$ is the velocity distribution function in which $M(t,t')$ act as the self-energy and vertex renormalizations as may also be seen in Fig.~\ref{Fig:SE} shows in its three lines that successively represent $M_{\rm SE}(t,t'), M_{\rm MT}(t,t')$ and $M_{\rm AL}(t,t')$. Since the vertex in  Fig.~\ref{Fig:SE} is a vector vertex, any scalar contribution coming across the particle-hole lines gives zero. This implies that $M_{\rm MT}(t,t')$, which is a ladder of particle-hole propagators connected across by 
the irreducible interaction $I_0(p,p')$ gives a zero contribution. 

The effective momentum independence of the self-energy leads to $[\hat{M}''_{\rm SE}(t,x)]_{LL'} \propto \delta_{L,L'}$ and we show in Appendix C that   
\begin{align}
\hat{M}''_{\rm SE}(t,x) &= 2 \bar{g}^2 N(0)T f(t) \delta(t-x) \hat{1}, 
\label{memory_SE_LCO}
\end{align}
where
\begin{align}
f(t) &= t \coth (t/2) .
\end{align}

The AL vertex corrections calculated in Appendix C is given as 
\begin{align}
\hat{M}''_{\rm AL}(t,x)&=
- \bar{g}^2 N(0)T [ F(t,x) + F(t,-x) ] \hat{B}. 
\label{memory_AT_LCO}
\end{align}
The effect of the vertex corrections 
is described by the matrix $\hat{B}$. The matrix elements of $\hat{B}$ are given in Appendix C in Eq. (\ref{vertex matrix}), and 
\be
F(t,x) &=& \frac{1}{4} 
\int_{-\infty}^{\infty}\frac{dy}{f(y)} \,{\rm sech} \frac{t-y}{2}  \,{\rm sech} \frac{y-x}{2}.
\label{funcF}
\ee

\subsection{Electrical and thermal conductivity} 

Let us define $\rho_{\rm SE}$ as an auxiliary quantity ignoring the vertex corrections, 
i.e., $\rho_{\rm SE}$ is the resistivity calculated with self-energy alone.
In Appendix C, it is derived that the true resistivity $\rho$ may be written in terms of lower and upper bounds as 
\be
\Big(\frac{\rho}{\rho_{\rm SE}}\Big)_{\rm L.B.} &=& \frac{1}{[\hat{C}^{-1}]_{11}}, 
\qquad
\Big(\frac{\rho}{\rho_{\rm SE}}\Big)_{\rm U.B.} = \frac{7 \zeta(3)/8}{[\hat{C}^{-1}]_{11} + 7 \zeta(3)/8 -1}.
\label{bounds}
\ee
Here 
\be
\label{hatC}
\hat{C} \equiv \hat{1} - \hat{B}.
\ee
Since $7 \zeta(3)/8 = 1.0518$, the bounds in Eq. (\ref{bounds})  are almost equal so that we can take the lower bound for evaluating the electrical resistivity.
It is also derived that $\sigma_{\rm SE} \equiv \rho^{-1}_{\rm SE}$, and 
\be
\sigma_{\rm SE} &=&
\frac{e^2\langle {\tilde v}_x^2 \rangle}{\bar{g}^2T} 
\int_{-\infty}^{\infty}  \frac{\tanh (t/2)}{4 t \cosh^2 (t/2) f(t) } dt
= \frac{7 \zeta(3)}{2 \pi^2} \frac{e^2\langle {\tilde v}_x^2 \rangle}{\bar{g}^2T}.
\label{result_sigmaSE}
\ee
In Sec. IV C, we explicitly derive the ratio $\rho/\rho_{\rm SE}$ for a circular Fermi surface in a square lattice because of Umklapp scattering.
For a small Fermi surface where Umklapp scattering is ineffective, the vertex corrections $\hat{B}$ can therefore never be ignored for $\sigma$ because $[\hat{C}^{-1}]_{11}$ diverges owing to the momentum conservation. Then the coefficient of the $T$ linear term in the resistivity vanishes. So Umklapp scattering is essential for the linear-in-$T$ resistivity. However, for the case of the critical fluctuations of the QXY-F model, it provides no temperature dependent factors. This is crucial to be in accord with experiments on single-particle self-energy and the  specific heat. 

The thermal conductivity is derived as 
\be
\kappa = 
\frac{\langle {\tilde v}_x^2 \rangle}{\bar{g}^2} 
\int_{-\infty}^{\infty}  \frac{t \tanh (t/2) }{4 \cosh^2 (t/2)} dt 
= \frac{\langle {\tilde v}_x^2 \rangle}{\bar{g}^2} .
\label{result_kappa}
\ee
Because of the Planckian dissipation, the thermal conductivity is a constant independent of temperature. 
The Lorentz number $L$ can then be described by
\be
L \equiv \frac{\kappa}{\sigma T} = \frac{\kappa}{\sigma_{\rm SE}T} 
\frac{\rho}{\rho_{\rm SE}}
= \frac{6}{7\zeta(3)} \frac{\rho}{\rho_{\rm SE}} L_0 ,
\ee
where $\rho = 1/\sigma$, $\rho_{\rm SE}=1/\sigma_{\rm SE}$, and $L_0 = (\pi^2/3) (k_{\rm B}/e)^2$ is the Lorentz number of normal metals. Since $\frac{6}{7\zeta(3)}=0.713$, the value of $L/L_0$ is about 70\% of the value of $\rho/\rho_{\rm SE}$.

To understand physically that Umklapp scattering is unnecessary for finite thermal conductivity, consider the single-particle decay of thermally excited fermions in the following process: 
A particle with a momentum ${\bf p}_1$ greater than the Fermi momentum $k_F$ interacts with a particle inside the Fermi sphere with a momentum ${\bf p}_2, (|p_2| < k_F)$ and decays into two particles with momenta ${\bf p}_3$ and ${\bf p}_4$ which are both outside the Fermi sphere. Since momentum is conserved in this process (${\bf p}_1+ {\bf p}_2 - {\bf p}_3 - {\bf p}_4 =0$), the charge current of the whole system does not decay because of the feedback effect of momentum coming back from other particles. Energy is also conserved in this process: If ${\epsilon}$ is the energy measured from the chemical potential, ${\epsilon}_1+{\epsilon}_2-{\epsilon}_3-{\epsilon}_4 = 0$. However, since ${\epsilon}_1, {\epsilon}_3$, and ${\epsilon}_4$ are positive while ${\epsilon}_2$ is negative, the product of ${\epsilon}$ and momentum is not conserved. For this reason, the energy current decays even without Umklapp scattering for the fermions, and the thermal conductivity is finite as a result of normal scattering.

These results can be put in context of the arguments attributed to Peierls, quoted for example in \cite{ziman2001electrons}.
Peierls has argued that Umklapp scattering is essential for finite thermal conductivity from phonons in the pure limit, but not for finite electronic thermal conductivity when that is defined, as is customary and done here, making sure that the electronic conductivity induced by thermal gradient is absent. We have substantiated this argument.
One should note however that the fermion self-energy $\Sigma$ on a lattice includes the effect of Umklapp scattering so that while one can say that the vertex renormalization is absent for the electronic thermal conductivity, it does not necessarily imply that the Umklapp scattering is absent.

It should be noted that there is no $\log(T)$ correction owing to mass renormalization for the electrical conductivity. There is also no $\log (T)$ enhancement of the thermal conductivity which appears in the critical specific heat because of the MFL single-particle quasiparticle renormalization. 

Recently, the breakdown of the Wiedemann-Franz law in strongly correlated electron systems and Weyl semimetals has been studied based on the Boltzmann equation \cite{Li-M2018, Zarenia-P-V2020}. We note that the disappearance of the quasiparticle renormalization in the thermal conductivity is generally true for other problems such as the heavy Fermi liquids, in which the self-energy is nearly momentum independent. 

\subsection{Seebeck coefficient} 

The thermoelectric power $S$, also called the Seebeck coefficient, can be given using the transport coefficients $L_{ij}$,
\be
S &=& \frac{L_{12}}{TL_{11}} .
\ee
In Sec. III, we have given a general expression of the off-diagonal transport coefficient $L_{12} = L_{21}$ at low temperatures [see Eq.~(\ref{L12general})]. Here we consider $L_{12}$ and thus $S$ for the QXY-F model, 
where $L_{11} = \sigma$ is given by Eq.~(\ref{sigma_LCO}).  The technical details of the calculations of the Seebeck coefficient follow the procedures given above for the electrical and thermal conductivity. We have collected them in Appendix E. Here we summarize the physics 
of some important differences from electrical and thermal conductivity of the results, which are very important in the experiments.

One notices from the results given in Appendix E 
that the renormalization factor $a$ does not appear in ${\tilde L_{12}}$ and in $L_{11}$. Therefore the Seebeck coefficient $S = a^{-1} \tilde L_{12}/L_{11}T$ is proportional to $a^{-1}$. In the case of the large Fermi surface as in the cuprates near quantum critical doping, Umklapp scattering is expected to make the vertex corrections ineffective for $L_{12}$ as well as for the electrical conductivity $L_{11}$. Then the first term can be dominant in Eq.~(\ref{tilde L21_v2}), leading to
\be
\frac{S}{T} &\simeq& 
-  \frac{2\pi^2}{7\zeta(3)} \frac{a^{-1}}{e}
\left( \frac{\langle {\tilde m}_{xx}^{-1} \rangle}{\langle {\tilde v}_x^2 \rangle} 
- \frac{1}{2}\frac{N'(0)}{N(0)} \right) . 
\ee
As found in experiments in the heavy fermion systems \cite{behnia2004thermoelectricity} and in cuprates \cite{Seebeck_LaNdSrCuO,Seebeck_Bi2201_PhysRevB.104.014515,Seebeck_cuprates} as well as the Fe-based compounds \cite{Seebeck_Fe_pnictides_PhysRevB.79.104504}, the magnitude of the Seebeck coefficient $S$ is proportional to $\propto a^{-1} T \propto T|\ln \frac{\omega_c}{T}|$, i.e. to electronic specific heat. $\omega_c$ is the upper cutoff in the fluctuation spectra and may be obtained by fit to measured $S$, which is often easier to measure than the specific heat. The sign of $S$ requires more detailed information about the electron dispersion, such as the inverse mass tensor and the energy dependence of the density of states. We remind again of the absence of the logarithmic mass enhancement in resistivity and in thermal conductivity.
Note also that unlike the case of thermal conductivity, there are no symmetry reasons that the vertex renormalization is absent in the Seebeck coefficient. Our result of the presence of the logarithmic mass enhancement in the Seebeck coefficient is for an ideal system without impurities, although a similar enhancement factor also appears as a result of impurity scattering~\cite{miyake2005theory}.

\begin{figure}[t]
\begin{center}
\includegraphics[width=0.8\columnwidth]{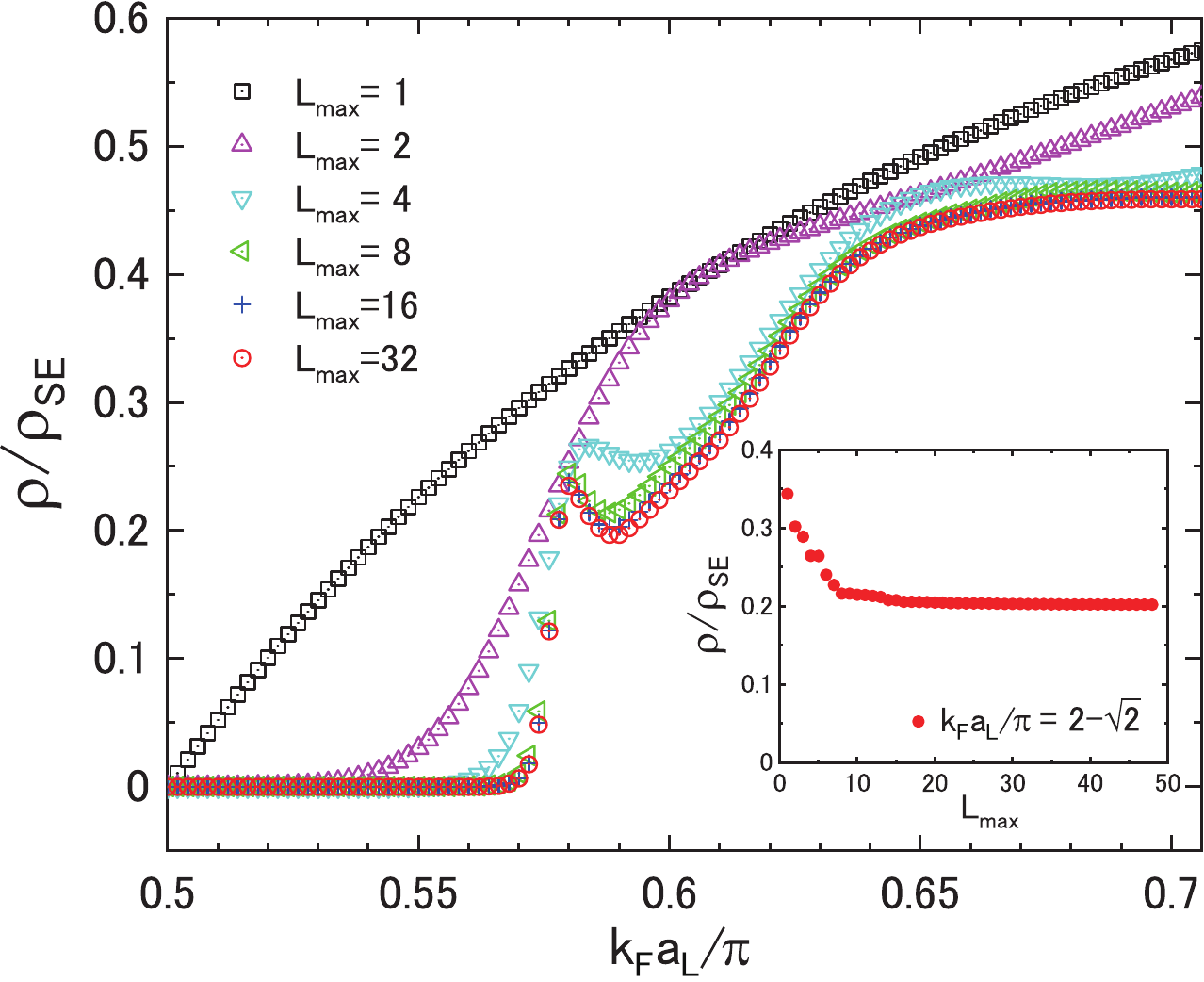}
\end{center}
\caption{Reduction factor of electrical resistivity owing to the vertex corrections, i.e. the Umklapp factor, for a circular Fermi surface of radius $k_F$ in a square lattice of unit cell ${a_L}$, calculated as a function of $k_F{a_L}/\pi$ ($1/2 < k_F{a_L}/\pi < 1/\sqrt{2})$ for different number of $L_{max}$. 
The inset shows the $L_{max}$ dependence of the Umklapp factor for $k_F{a_L}/\pi = 2-\sqrt{2}$.}
\label{Fig:U1}
\end{figure}
\begin{figure}[t]
\begin{center}
\includegraphics[width=0.7\columnwidth]{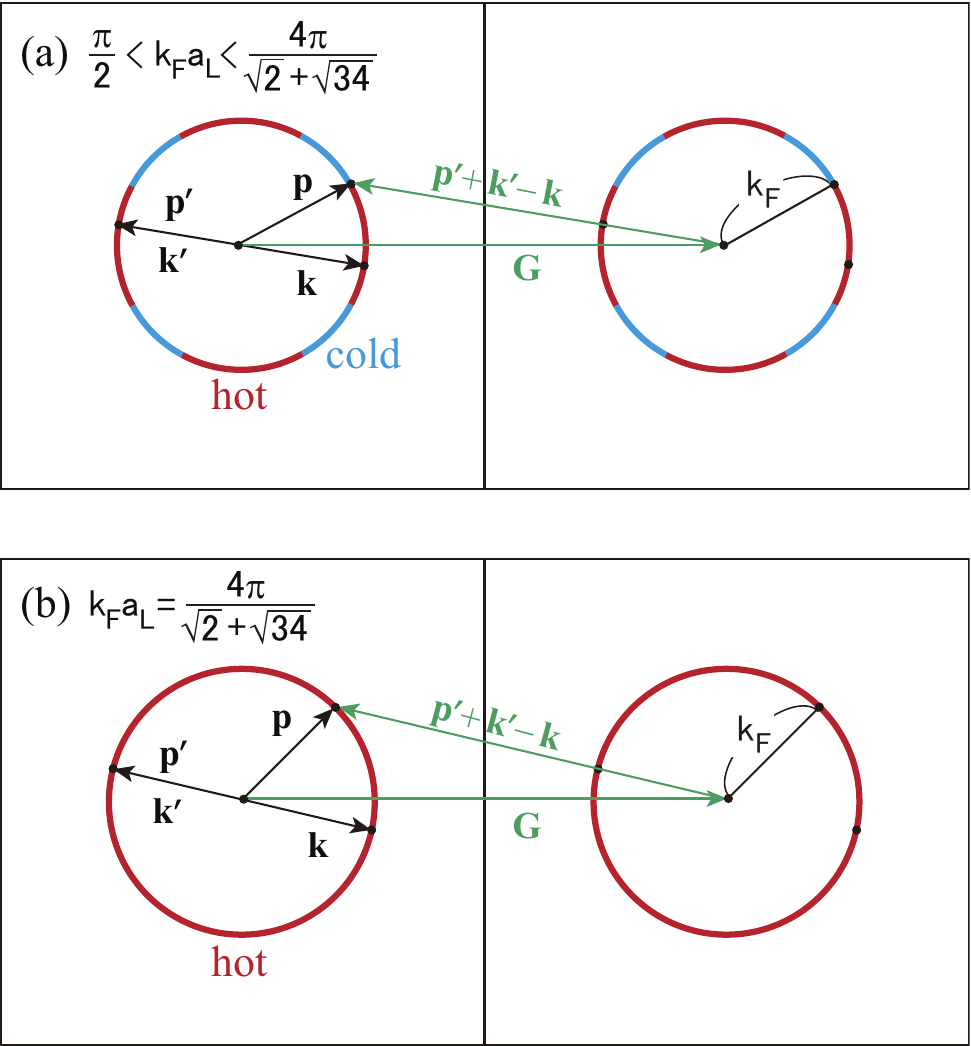}
\end{center}
\caption{Kinematics of Umklapp scattering for ${\bf p}-{\bf p}'+{\bf k}-{\bf k}'={\bf G}$ with ${\bf p}'={\bf k}'=-{\bf k}$ and ${\bf G}=(2\pi/{a_L},0)$. Two Brillouin zones are considered and it is demonstrated that for $k_F{a_L}/\pi < 0.5$, their is no Umklapp scattering. 
(a) There is a "hot" part of the Fermi surface near the corners where Umklapp is allowed but the major part of the Fermi surface is "cold" for  $0.5<k_F{a_L}/\pi<4/(\sqrt{2}+\sqrt{34}) \approx 0.552$. (b) For larger $k_F{a_L}$, the entire Fermi surface is hot and only then is there finite resistivity. }
\label{Fig:U-2}
\end{figure}

\subsection{Results for Umklapp vertex on a square lattice with a circular Fermi surface} 

The reduction factor $\rho/\rho_{\rm SE}$ of electrical resistivity owing to the vertex corrections is 
called the Umklapp factor, 
where $\rho$ is the actual resistivity and $\rho_{\rm SE}$ is calculated by putting $\hat{B}=\hat{0}$. As shown in Appendix C, the Umklapp factor is given by the inverse of the 11th element of the inverse matrix of $\hat{C}$. 
For a circular Fermi surface described by ${\tilde{\epsilon}}_{{\bf p}} \!=\! (p^2 - k_F^2)/2{\tilde m}=0$, from Eqs.~(\ref{result_sigmaSE}) and (\ref{result_kappa}), 
$\rho_{\rm SE}=\frac{2\pi^2}{7\zeta(3)}\frac{(\bar{g}{\tilde m})^2}{\pi n e^2}T$ and the $T$-independent thermal conductivity is given by 
$\kappa=\pi n/(\bar{g}{\tilde m})^2$ with the electron number density $n$ and the electron unrenormalized mass ${\tilde m}$. The evaluation of $\hat{C} = \hat{1} - \hat{B}$ is, however, quite nontrivial even for a circular Fermi surface and given in Appendix D. The infinite-dimensional matrix $\hat{C}$ can be approximated by a square matrix of $L_{max} \times L_{max}$ and the limit of $L_{max} \to \infty$ gives the correct value.  
We present the results for numerical evaluation for the Umklapp factor $\rho/\rho_{\rm SE}$ in Fig.~\ref{Fig:U1} for various $L_{max}$ ($=1,2,4,8,16,32$) and for a circular Fermi surface of radius $k_F$ in a square lattice of unit cell ${a_L}$.  

We first turn to the  kinematics for various ratios $k_F{a_L}/\pi$, which underlies the calculations. This is explained in Fig.~\ref{Fig:U-2}. For $k_F{a_L}/\pi<0.5$, where the diameter of the Fermi surface is less than half of the reciprocal lattice vector, the resistivity is zero because there is no Umklapp scattering. When $0.5<k_F{a_L}/\pi<4/(\sqrt{2}+\sqrt{34}) \approx 0.552$, the momentum involved in Umklapp scattering is restricted to the four "hot" portions of the Fermi surface shown in red, and only normal scattering occurs in the "cold" portions shown in blue in Fig.~\ref{Fig:U-2}(a). The entire Fermi surface becomes "hot" for $k_F{a_L}/\pi \approx 0.552$ as shown in  Fig.~\ref{Fig:U-2}(b). We find in Fig.~\ref{Fig:U1} that the resistivity is zero if there is any part of the Fermi surface that is "cold". This is an interesting effect arising from the fact that $\sigma$ is proportional to  $[\hat{C}^{-1}]_{11}$ in Eq. (\ref{resistivity_ratio}), and $\hat{C}$ has obviously zero eigenvalues. As mentioned in Appendix D, we can write $\hat{C} = \hat{C}_N + \hat{C}_U$ in general, where $\hat{C}_N$ and $\hat{C}_U$ are the contributions from normal and Umklapp scatterings, respectively. ${\hat C}_U$ is not a zero matrix in the presence of the "hot" part, but it has an infinite number of zero eigenvalues in the presence of the "cold" part. This is because the "cold" regions always short-circuit the "hot" regions, or more precisely because we can choose $\theta_{\bf p}$ as a conserved quantity in 
Eq.~(\ref{Ward--Takahashi2}), which has a value only at points in the "cold" regions and is always zero in the "hot" regions. In the present two dimensions, $\hat{C}_N$ is a zero matrix, i.e., $\hat{C}={\hat C}_U$, so this choice of $\theta_{\bf p}$ does not produce any extra contribution from normal scattering. As a result, there are an infinite number of conserved quantities for a two-dimensional Fermi surface with the "cold" part. This anomaly disappears when impurity scattering is taken into account, because then no region is "cold" to begin with. 

\begin{figure}[t]
\begin{center}
\includegraphics[width=0.8\columnwidth]{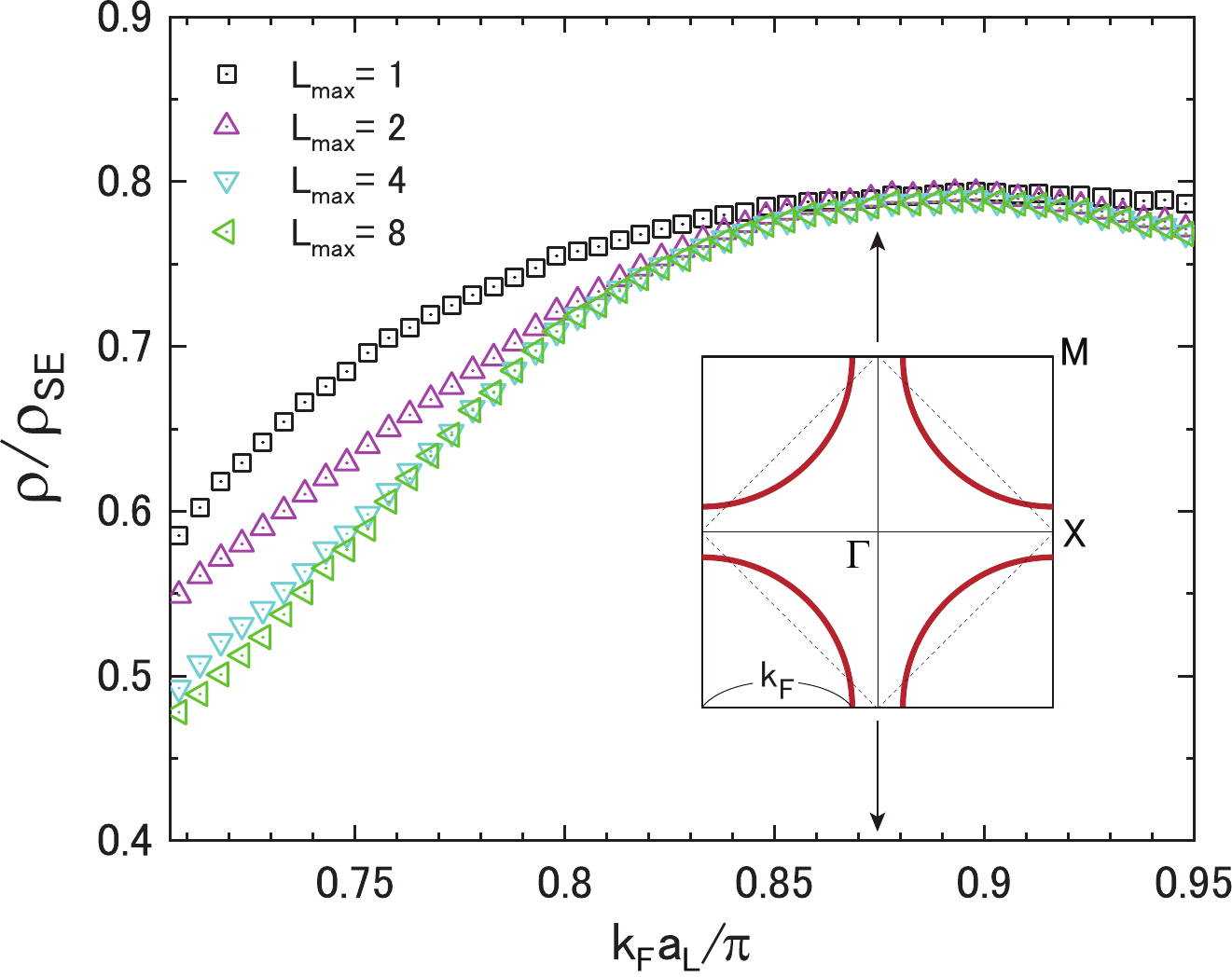}
\end{center}
\caption{Umklapp factor of the resistivity for a circular Fermi surface on a square lattice with $k_F{a_L}/\pi > 1/\sqrt{2}$, where  higher Umklapp vectors such as $(2\pi/{a_L}, 2\pi/{a_L})$ become players. 
The inset shows a circular Fermi surface of radius $k_F \approx 0.874 \pi/{a_L}$ centered at the M point, which corresponds to about 20\% hole doping. 
Around this value of $k_F$, the convergence with respect to $L_{max}$ is fast and the Umklapp factor is nearly constant at about $0.78$.}
\label{Fig:U2}
\end{figure}

In Fig.~\ref{Fig:U1}, we find that the convergence with respect to $L_{max}$ is slow. This is especially so around $k_F{a_L}/\pi=2-\sqrt{2}\approx0.586$, where $\rho/\rho_{\rm SE}$ has a local minimum. But we have investigated this region more exhaustively with increased $L_{max}$ to assure that the value of the Umklapp factor there is about 0.2 but saturating at about 0.5 at large $k_F{a_L}$. If the Fermi surface is large enough, so that higher Umklapp vectors $(2\pi/{a_L}, 2\pi/{a_L})$ become players and the true resistivity $\rho$ then becomes closer to $\rho_{\rm SE}$. For the about 20\% hole doped cuprates near quantum-critical doping, the Fermi surface is large. We have also calculated $\rho/\rho_{\rm SE}$ for $k_F{a_L}/\pi>1/\sqrt{2}$, where there is Umklapp scattering also for the next reciprocal vectors. As seen in Fig.~\ref{Fig:U2}, the convergence with respect to $L_{max}$ is fast and the Umklapp factor is nearly constant at about $0.78$ around $k_F{a_L}/\pi = 2\sqrt{3/5\pi} \approx 0.874$, which corresponds to the radius of an oblate Fermi surface centered at the M point of the 20\% hole doped cuprates when approximated by a circle. We would therefore expect $0.5 < \rho/\rho_{\rm SE} < 1$. This is in line with the estimates given in \cite{Varma-rmp2020} for the relative magnitude of the imaginary part of the self-energy in the diagonal direction, the coefficient of the $T \ln T$ specific heat and the resistivity scattering rate estimated from experiments. 

In this context, one of the results in \cite{MFII} is worth recalling; for $d>2$ where ${\hat C}_N$ is not a zero matrix, normal scattering also contributes to the electrical resistivity once Umklapp scattering is present. 
For transport theory in dimension $d \to \infty$ or for infinite number of neighbors, as in dynamical mean-field calculations, the Umklapp factor is identity, implying that the normal scattering contributes to the electrical resistivity as well as the Umklapp scattering. This should be equally true for the SYK-type models \cite{SYK2021} in which $N$ sites are coupled to each other with $N \to \infty$. 

\section{Results for conductivity from Boltzmann equation} 

Equation~ (\ref{Kubo-1}) in the dc limit is the familiar Boltzmann equation for conductivity with $\Phi_{\bf p}$ is the velocity distribution function. It is well know that $\Phi_{\bf p}$ obeys the integral equation, 
\be
\label{Phip}
\Phi_{\bf p} = \sum_{\bf p'} \mathcal{T}_{{\bf p, p}'} \Big(- \frac{\partial f}{\partial {\bf \epsilon_{\bf p'}}}\Big) \tilde{{\bf v}}_{\bf p'},
\ee
where $\mathcal{T}_{{\bf p, p}'}$ is the collision operator,
\be
\label{Tpp}
\mathcal{T}_{{\bf p, p}'} = \tau_{\bf p}~ \delta({\bf p, p}') + \sum_{\bf p'} {\mathcal C}_{\bf p, p"} ~\mathcal{T}_{{\bf p", p'}}.
\ee
The first term is the ``scattering out" term
\be
\label{taup}
\frac{1}{\tau_{\bf p}} = \sum_{\bf p'} {\mathcal C}_{\bf p, p'},
\ee
while the second is the ``scattering in" term. 

Everything above in this section is familiar and formal. The necessary microscopic calculations for fermion scattering in this paper, which are an evaluation in effect of $\Phi _{\bf p}$, have been for coupling of fermions to fluctuations of a quantum-critical point of an XY model but the calculations are done only for a circular Fermi surface so that $\frac{1}{\tau_{\bf p}}$ is angle independent.   For the fluctuations of the quantum XY model and their coupling to fermion for Fermi surface which is not circular  the  self-energy, as calculated in Appendix B, is actually (weakly) angle dependent. This is in fact what is realistic. In experiments, the resistivity and other anomalous transport properties  have the same temperature dependence as calculated here for a circular Fermi surface. One cannot compare the coefficients in the experiments with what is calculated except for the order of magnitude. But it is important to show why the temperature dependence in experiments are the same as for the circular Fermi surface. We do this below based on general properties of the velocity distribution function $\Phi_{\bf p}$ following from the nature of the fluctuations giving ${\mathcal C}_{\bf p, p'}$.  

\subsection{Temperature dependence of conductivity for a non-circular Fermi surface} 

Using Eqs. (\ref{Phip}) and (\ref{Tpp}), it follows that
\be
\label{Phip1}
\Phi_{\bf p} = \tau_{\bf p}
\Big( -\frac{\partial f}{\partial \epsilon_{\bf p}}} {\bf {\tilde{v}}_{\bf p}  - \sum_{\bf p'}  {\mathcal C}_{\bf p, p'} \Phi_{\bf p'}\Big).
\ee
As noted above  $\tau^{-1}_{\bf p}$ is the single-particle scattering rate or the imaginary part of the self-energy given by Eq. (\ref{taup}). It is shown in Appendix A for a general fourfold asymmetric Fermi surface to be
\be
\frac{1}{\tau_{{\bf p}}} =  \overline{g}^2 t T \coth \big(\frac{t}{2}\big) ~ \mathcal{F}({\bf p}), 
\ee 
where  $\mathcal{F}({\bf p})$ is a temperature-independent dimensionless function with the symmetries of the square lattice. But the same calculation gives us the collision operator through Eq. (\ref{Sig1}),
\be
\label{collop}
{\mathcal C}_{\bf p, p'} &=& - \overline{g}^2 t T \coth \frac{t}{2} {\mathcal F}({\bf p, p'}), \\
{\mathcal F}({\bf p, p'}) &=&  \frac{ ({\bf p} \times {\bf p}')^2}{({\bf p} - {\bf p'})^2 + \xi_r^{-2}} 
\delta(\tilde{\epsilon}_{{\bf p'}})
\ee
This is a product of a function of energy and  temperature and a function of the momentum variables. Moreover the temperature dependence of $\tau_{\bf p}$ is identical to that of ${\mathcal C}_{\bf p, p'}$. 
Define a  function
\be
M({\bf p}) \equiv \frac{\Phi_{\bf p}}{\overline{g}^2 t T \coth \frac{t}{2}}
\ee
and inserting on both sides of Eq. (\ref{Phip1}), one directly finds an integral Equation for $M({\bf p})$ in which all terms are temperature and energy independent. It then follows that $\Phi({\bf p})$ has the same temperature and energy dependence for all ${\bf p}$.

Inserting this result in Eq. (\ref{Kubo-1}), it follows that the conductivity for fermions scattering from the fluctuations of the quantum XY model is linear-in-$T$ for a general Fermi surface. We cannot calculate the coefficient for such a Fermi surface. As seen above, such a calculation is quite elaborate even for a circular Fermi surface.  In deriving that the resistivity is $\propto T$ for a general Fermi-surface, we have used the fact that the single-particle self-energy is $\propto T$ for all harmonics of the square lattice, which itself follows from the product of a function of $\omega/T$ and a function of momentum of the critical fluctuations and their small spatial correlation length.

\section{Conclusions} 

It has been apparent through dogged work by hundreds of theorists that the remarkably simple and universal properties near quantum criticality of a large number of quasi-two-dimensional systems may only be explained through the $\omega/T$ scaling of critical fluctuations proposed in the MFL model in 1989. The only physically realizable model in which such a spectrum has been derived is the QXY-F model. The  frequency and temperature dependence are identical to the MFL  phenomenology, but there is a momentum dependence both in the fluctuations and in their coupling to fermions. But because of the unusual form of the fluctuation spectrum, never encountered in the storied history of critical phenomena, the functional form of the fermion self-energy is shown not to change, in the necessary physical regions, from the phenomenology.  However, even for a circular Fermi surface, although some properties like the single-particle spectrum and the specific heat are very easy to calculate, the calculation of the transport properties is subtle and, to be completely convincing, requires a technical and detailed calculation. This is accomplished in this paper in the simplest manner we know. We have also emphasized the (already know) physical principles and conservation laws underlying transport properties generally.

We have solved the integral equation for the vertex coupling to external fields in the Kubo expression for electrical and thermal conductivities and the Seebeck coefficient, using scattering of fermions from the known propagator for the quantum fluctuations of the QXY-F model. We have derived numerically exact results for a circular Fermi surface at low temperatures for transport properties. For more general Fermi surfaces, qualitatively identical results are shown. All of this has been possible only because of the simplicity and the unusual nature of the scale-invariant spectra of fluctuations. This is also the only example of a physically applicable model we know for which the Kubo equations for transport have been solved.

The model solved for its transport and thermodynamics has direct applicability to a variety of physical problems of experimental interest that give experimental results with  linear-in-$T$ and linear-in-$H$ resistivity, temperature-independent thermal conductivity and $T \ln (\omega_c/T)$ specific heat, with coefficients that are simply related to each other.  The general theory  presented should be of considerable technical interest. Among the several new results derived here are the presence of the logarithmic-in-temperature mass enhancement in the leading term in the Seebeck coefficient and its absence in the thermal conductivity. There is a vast list of experiments that are explained unambiguously through this work but the absence of the mass enhancement of the specific heat in the thermal conductivity is yet to be tested.
Although all calculations in this paper are for an ideal system without impurities, it is an interesting and important question what impurities do to quantum criticality~\cite{m2v_PhysRevLett.95.207207,m2v_PhysRevLett.88.226403, VARMA20251354703}. 

The problem of "strange metals" in the quantum-critical region, which are all marginal Fermi liquids if the resistivity is linear-in-$T$ and the specific heat or thermopower is $T \ln(\omega_c/T)$, is a truly interesting quantum phenomenon with a wide universality class. There has been unfortunately much fuzzy thinking on this problem, with models applicable as well as inapplicable. We hope that this work, through its precise and controlled results and their  agreement with detailed experiments, provides an unambiguous understanding  of this class of anomalies near quantum-critical points discovered in experiments in a variety of different materials with quite different order parameters. These include the cuprates with loop-current order without breaking translation symmetry,  planar antiferromagnets and incommensurate Ising antiferromagnets, TBG if it indeed has loop-current order breaking translation symmetry, as well as TBWSe, which is a classic case of a two-dimensional XY model \cite{ChubukovV2025}. We have emphasized that the reason for the universality near quantum criticality is that they are all described by the QXY-F model.  In some antiferromagnetic quantum-critical points, questions of cross over to this model are well posed and have been addressed but only based on scaling concepts \cite{VARMA20251354703}.  It follows also that the superconductivity in this universality class is promoted by scattering of fermions from the same fluctuations that give the normal state strange metal anomalies. This has been explicitly tested by inversion of ARPES results in cupates \cite{Bok_ScienceADV}.\\

\acknowledgements
HM is grateful to M. Ogata and H. Matsuura for fruitful discussions. 
CMV wishes to acknowledge several discussions and email exchanges with Dominic Else and Senthil Todadri which were useful in understanding their work. The work of HM was partly supported by Grants-in-Aid for Scientific Research from the Japan Society for the Promotion of Science (Grant No. JP21K03426) and JST-Mirai Program, Japan (Grant No. JPMJMI19A1). 
CMV wishes to thank James Analytis, Robert Birgeneau and Joel Moore for arranging for him to be a "re-called Professor" at the Physics department of University of California, 
Berkeley, where part of this work was done and to Aspen Center for Physics where part of this work was done last summer. Aspen Center for Physics is partially supported by the National Science Foundation of USA through Grant PHY-2210452.

\appendix

\section{Summary of the critical fluctuations of quantum XY model \\coupled to fermions} 

\subsection{The model}
The quantum XY model in (2+1) dimensions is defined in terms of the action $S_{qxy}$ of a rotor of fixed length and angle $\theta({\bf x}, \tau)$ at a point ${\bf x}$ and imaginary time $\tau$, which is periodic in the inverse temperature $(0, \beta)$, is given by the following action:
\begin{eqnarray}
\label{modelxy}
S_{qxy} &=&-K_0 \sum_{\langle {\bf x, x}' \rangle} \int_0^{\beta} d \tau \cos(\theta_{{\bf x}, \tau} - \theta_{{\bf x}', \tau}) \nonumber \\
& +& \frac 1 {2E_c} \sum_{{\bf x}} \int_0^\beta d \tau \left( \frac{d \theta_{{\bf x}}}{d\tau}\right)^2 .
\label{eq:model}
\end{eqnarray}
The first term is the potential energy of interactions and the second term is the kinetic energy owing to the angular momentum of the rotors. The variable $\theta$ refers to different quantities in different physical situation. It refers to the direction of the anapole vector in the loop-current order in cuprates \cite{simon-cmv}, to the in-plane antiferromagnetic-order vector in some heavy fermion, transformed on a bipartite lattice into a model coupled ferromagnetically in the plane \cite{cmv_PhysRevLett.115.186405,CMV-IOP-REV}, or to the phase of an incommensurate antiferromagnetic Ising order in another heavy fermion \cite{cmv_PhysRevLett.115.186405,CMV-IOP-REV}. Lattice anisotropy if of symmetry fourfold or more is irrelevant in the quantum model and is ignored. Recently, it has been suggested \cite{Fu_V2023} that the loop-current order proposed in Moir{\'e} TBG \cite{Zaletel2020, Berg2021}, (which is remarkably similar to an order presented for graphene because of nearest-neighbor interactions \cite{Zhu-A-V2013}), and proposed in TBWSe also fall in the quantum-XY class.
  
The model of Eq.~(\ref{modelxy}) is equivalent in the limit $T \to 0$ to that of an XY model in three dimensions. It has Lorentz-invariant fluctuations at long wavelengths, with a propagator,
\begin{eqnarray}
\label{D0}
D_0(q, \omega) = \frac{2 c q}{\omega^2 - c^2 q^2 + i\omega \delta},
\end{eqnarray}
where $\delta$ is infinitesimal. This is quite inadequate to address questions in relation to problems of our interest where such collective modes  come from  interaction among the fermions, and a residual interaction of the collective variables to fermions is inescapable and changes the fluctuation spectrum in an essential way. In the present problem, it changes the dominant excitations to topological excitations in space and in time \cite{Aji-V-qcf1, ZhuChenCMV2015, ZhuHouV2016, Hou-CMV-RG}, as summarized in Appendix A 3 below. 

To generate the contribution of the effective action $S_{c-f}$ for the collective fluctuations caused by the coupling to the fermions, we must first find the coupling of fermions to the collective variables.
The appropriate coupling to fermions must be found from the same interactions among the fermions that lead to the collective fluctuations.  A summary of such couplings is derived in Appendix C 4. The variable $\theta({\bf x}, \tau)$ is not gauge invariant and cannot couple to any physical variable that can be constructed from fermion operators.  On the other hand $\nabla \theta({\bf x}, \tau)$ is proportional to a collective mode current and if $\theta$ refers to the direction of a vector that is time-reversal and inversion-odd, it  couples to the fermion current, (spin-current for the case of XY antiferromagnets or ferromagnets). Similarly, the conjugate variable in the quantum-rotor problem, i.e. the angular momentum $\frac{d \theta ({\bf x}, \tau)}{d\tau}$ which is proportional to the orbital magnetic moment  or in a Hamiltonian formulation ${\bf L}_z({\bf x}, \tau) = i \frac{\partial}{\partial {\hat{{\bf z}}}}$ similarly couples to the local fermion angular momentum or orbital moment. 

To generate the effective action $S_{c-f}$ for the collective fluctuations owing to the coupling to the fermions, we must integrate over the fermion 
current correlations. These are correlations of unconserved quantities. The imaginary part of the correlation function for $T \to 0$ and the limit $\omega \to 0$ after $q \to 0$ is given by $|\omega| \sigma(T=0)$, where $\sigma$ is the current or angular momentum conductivity. These are finite in the limit $T \to 0$ only because of defects and we take them to be a constant $\alpha$ normalized to $e^2/h$. This leads to the form $S_{c-f}$ given in Eq. (\ref{scfqw}) below. 
This form is of the Caldeira-Leggett \cite{CaldeiraLeggett} type, although the physical basis is quite different.  It is given in momentum and frequency space by
\be
\label{scfqw}
S_{c-f} = \sum_{{\bf q}, \omega_n} \frac{\alpha}{4\pi^2} |\omega_n| q^2 |\theta({\bf q}, \omega_n)|^2.
\ee 
Here $\alpha$ is the  conductivity of the fermions in the limit $q \to 0, \omega \to 0, T \to 0$ made dimensionless in terms of $e^2/h$. 
This looks much more formidable when transformed to space and imaginary time, which is necessary to do in order to use the quantum Monte Carlo technique.
\be
\label{c-f}
S_{c-f} &=& \frac{\alpha}{4\pi^2} \sum_{\langle{\bf x, x}'\rangle} \int d \tau  d\tau' \frac {\pi^2}{\beta^2} 
\frac {\left[(\theta_{{\bf x}, \tau} - \theta_{{\bf x}', \tau})  -(\theta_{{\bf x}, \tau'} - \theta_{{\bf x}', \tau'}) \right]^2}{
\sin^2\left(\frac {\pi |\tau-\tau'|}{\beta}\right)} .
\ee

\subsection{Summary of results from quantum Monte Carlo}

\begin{figure}[h]
\begin{center}
\includegraphics[width=0.95\textwidth]{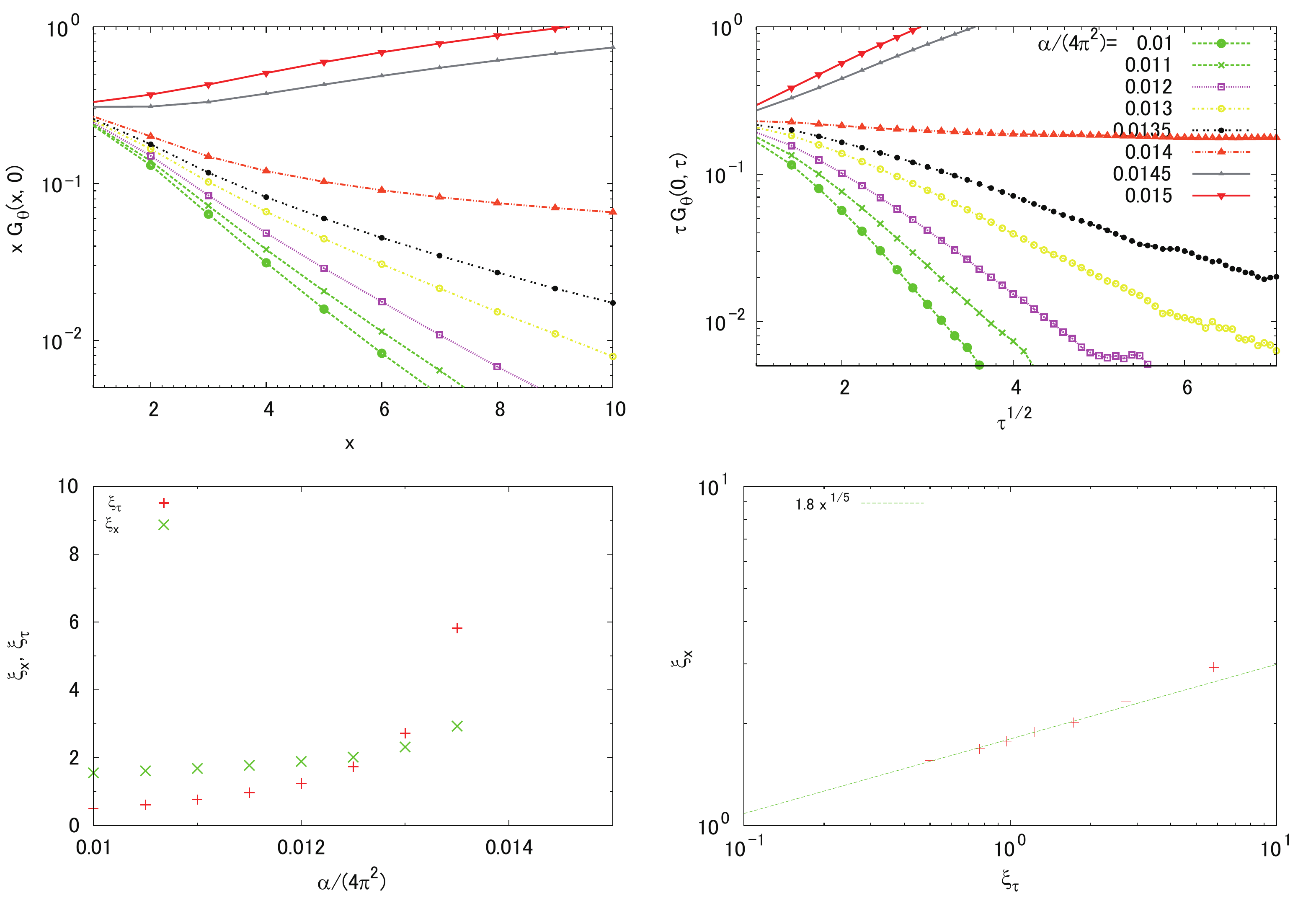}
\end{center}
\caption{This figure gives an example of the correlation function $D({\bf x}, \tau)$ (called $G$ in the original paper), calculated in \cite{ZhuHouV2016}. All results are for fixed $K_0$ and $E_c$ and varying $\alpha/4\pi^2$. Top left at small $\tau$ $D$ is shown as a function of $x$, while top-right shows it at small $x$ as a function of $\tau$. With such fixed values, there is a critical point for $\alpha/4\pi^2 = 0.014 \pm .0002$. These curves can be collapsed and shown to be of a scaling form near the critical form from which the correlation length in time $\xi_{\tau}$ and in space $\xi_x$ can be deduced as shown at bottom left. In bottom right, the relation between $\xi_{\tau}$ and $\xi_x$ is shown to be $\xi_x \sim \xi_\tau^{1/z}$. The minimum value of $z$ with which these results can be fit is $6$ but $z \to \infty$, i.e. $\xi_x \sim \log \xi_{\tau}$ fits very well.}
\label{Fig:Flucs}
\end{figure}

The QXY-F model which has been investigated by quantum Monte Carlo \cite{ZhuChenCMV2015, ZhuHouV2016} very extensively is 
\be
\label{qxyF}
S_{qxy-F} \equiv S_{qxy} + S_{c-f}.
\ee
As a function of $K_0, E_c$ and $\alpha$ it has three different lines of transitions. We will be interested in the problem under discussion where the quantum-critical point fluctuations are at the transition in which $\cos(\theta({\bf x}, \tau))$ orders both in space and time. So the critical modes are given by the correlation function of the fluctuations 
\be
\label{D}
D({\bf r}, \tau) &\equiv& <e^{-i \theta({\bf r}, \tau)} e^{i \theta(0, 0)} > 
~\propto ~ <L_z({\bf r}, \tau) L_z(0,0)> .
\ee
The proportionality of the fluctuations of the angular momentum variable to those of $e^{i \theta}$ has been shown in an appendix in \cite{Aji-V-qcf2}. (Actually, the derivation there is for momentum independent fluctuations. But it is trivial to extend it using the separable form of the fluctuations to momentum-dependent fluctuations.) The result for the fluctuations are very accurately given by \cite{ZhuChenCMV2015, ZhuHouV2016, Stiansen-PRB2012} 
\be
\label{spectra}
D({\bf r}, \tau)  = D_0 \frac{\tau_c^2}{\tau} ~e^{-(\tau/\xi_{\tau})^{1/2}}~\ln (r/a_L) e^{- r/\xi_r}.
\ee
$D_0$ provides the magnitude of the fluctuations that is given by the square magnitude of the orbital current moments per unit cell. $\tau_c^{-1}$ is the ultraviolet energy cutoff and $a_L$ is the lattice constant. Further
the spatial correlation $\xi_r$ is negligible for all practical purposes compared to the 
temporal correlation length,
\be
\label{corrlengths}
\xi_r/a_L = \ln (\xi_{\tau}/\tau_c).
\ee
The most remarkable feature of (\ref{spectra}) is that they are of product form in time and space. This arises because of the nature of the two kinds of orthogonal topological excitations responsible for the correlation functions noted in \cite{Aji-V-qcf1, Aji-V-qcf3} one of which propagates only in space and the other only in time. 
This becomes more evident when one expresses the action in a different form, summarized below in Appendix A-3. The orthogonality comes from the fact that one of the excitations, the vortices are proportional to curl of the velocity field and the other the warps to the (time derivative of) the divergence of the velocity field; see next subsection. The divergence is present only because of coupling to fermions.

An example of the accuracy of the extensive calculations \cite{ZhuChenCMV2015, ZhuHouV2016} from which the above conclusions are drawn is given by Fig.~\ref{Fig:Flucs}. Many more results with various different parameters may be found in these references.

\subsection{Analytic RG calculations on the model} 

After making the Villain transformation,
\be
\cos(\theta_{{\bf x}, \tau} - \theta_{{\bf x}', \tau}) \to \sum_{\bf {m}({\bf x,x'}, \tau)}  (\theta_{{\bf x}, \tau} - \theta_{{\bf x}', \tau} -  \bf {m}({\bf x -x'}, \tau))^2, 
\ee
where ${\bf {m}}({\bf x,x'}, \tau)$ is a space and time-dependent integer field,
 the model given by Eq.~(\ref{qxyF}) for $S_{qxy-F}$ is harmonic in $\theta$. Integrating over such spin-wave like fluctuations, as is done for the classical XY model to generate an action in terms purely of vortices, leaves a model for topological excitations in space and in time \cite{Aji-V-qcf1, Aji-V-qcf3, Hou-CMV-RG}. A  choice of the degrees of freedom exactly transforms the resulting Lagrangian into a remarkably simple form and provides crucial insight to the Monte Carlo results. The choice consists in defining two varieties of topological charges through the effective integer velocity field ${\bf m}({\bf r}, \tau)$ on sites of the space-time lattice. ${\bf m}({\bf r}, \tau)$ has both a transverse and a longitudinal part
\be
{\bf m}({\bf r}, \tau) &=& {\bf m}_t({\bf r}, \tau) + {\bf m}_{\ell}({\bf r}, \tau), \\ 
\nabla \cdot {\bf m}_{t} &=& 0, ~ \nabla \times {\bf m}_{\ell} = 0.
\ee
Quantized topological charges for vortices $\rho_v$ and for warps $\rho_w$ are defined by
\be
\nabla \times {\bf m}_t({\bf r}, \tau) &=&\rho_v ({\bf r}, \tau), \\
\frac{d}{d\tau} \big(\nabla \cdot {\bf m}_{\ell}\big) &=& \rho_w({\bf r}, \tau).
\ee
$\rho_v$ and $\rho_w$ are orthogonal fields. In terms of them
 $S_{qxy-F}$ can be rewritten as
\be
S_{qxy-F} &=& S_v + S_w + S_w',\\
S_v &=&  \frac{J}{2\pi}\sum_{i\ne j}\rho_v({r}_i,\tau_i)\ln\frac{|{r}_i-{r}_j|}{a_c}\rho_v({r}_j,\tau_i) 
+\ln y_v\sum_{i}|\rho_v({r}_i,\tau_i)|^2,\\  
S_w &=& \alpha \sum_{i\ne j}\rho_w({r}_i,\tau_i)\ln\frac{|\tau_i-\tau_j|}{\tau_c}\rho_w({r}_i,\tau_j) 
+ \ln y_w\sum_{i}|\rho_w({r}_i,\tau_i)|^2,\\
S_w' &=& g\sum_{i\ne j}\rho_w({r}_i,\tau_i)\frac{1}{\sqrt{|{r}_i-{r}_j|^2+v^2(\tau_i-\tau_j)^2}}
\rho_w({r}_j,\tau_j).
\ee
$S_v$ is the action of vortices that interact logarithmically in space but is local in time. $S_w$ is the action of warps that interact logarithmically in time but is local in space.  $y_v$ and the $y_w$ are the fugacities of the vortices and warps respectively. The last term $S_w'$ effectively couples the vortices and warps through $g$ and $v$ which flow and control which of the two dominates the correlations to determine the correlation functions. The flow of $v^2$ under renormalization is equivalent to the flow of the dynamical critical exponent $z$. In most of the range of parameters, warps dominate the order and force the order of vortices. Only results in this range have been quoted here. Direct evidence for warps and vortices  is observed in the quantum Monte Carlo calculations and their correlations can be related to the fluctuations given by the order-parameter fluctuations defined by Eq. (\ref{D})

The transformed problem in terms of two kinds of topological excitations is as well soluble as the Kosterlitz solution \cite{KT1973, JKKN1977} of the classical XY model in first-order RG. The first-order RG solution reproduces most but not all features of the Monte Carlo solution. Two differences are noteworthy. In the RG calculations, the critical point of interest occurs for $\alpha =1$, regardless of the other flowing parameters in RG. In the Monte Carlo calculations, there is a critical surface in the $K_0, E_c, \alpha$ plane in which the exponents are identical. 
Moreover a remarkable result from Monte Carlo is that at the critical surface, not only are the correlation functions of the same form but the amplitude $D_0$ is also independent of them. This is not reproduced by the one loop RG, nor is the fact that the correlation decays as $e^{- (\tau/\xi_\tau)^{1/2}}$ and not as  $e^{- (\tau/\xi_\tau)}$. 

The correlation function calculated by RG can be continued to real frequency and momentum only numerically for finite correlation lengths.  The  results can be well fitted to the form given by the analytical calculations, 
\be
\label{Dqw}
D({\bf q}, \omega) &\approx & D_0 \frac{\xi_r^{-2}}{q^2 + \xi_r^{-2}}
\Big(\ln  \big|\frac{\omega_c}{max(\omega, \pi T , \xi_\tau^{-1})}\big|  - i \tanh \frac{\omega}{\sqrt{(2T)^2 + \xi_{\tau}^{-2}}}\Big)  .
\ee

\subsection{Derivation of coupling of collective fluctuations to fermions and of $S_{cf}$}

The coupling to loop-currents has been derived in Ref. {\cite{ASV2010}.  Here we  derive them in a simpler way. Let us consider a partially filled band in the loop-current state in a copper-oxide lattice, whose creation and annihilation operators are denoted by $c_{\bf k}^+, c_{\bf k}$ and has eigenvalues 
from Eq. (\ref{H0}) which are 
\be
\label{H0}
H_0 = \epsilon \left(\frac{k_xa}{2} + \phi_x, \frac{k_ya}{2} + \phi_y \right)c_{\bf k}^+ c_{\bf k}.
\ee
 $(\phi_x, \phi_y)^T$, the phase difference across the lattice constant  in the x and y directions in a unit cell, specify the loop-current order parameter. For the fluctuations, we must consider $(\phi_x, \phi_y)({\bf r})$ which are space dependent. It is straightforward to cast the results below to fluctuations of an XY ferromagnetic or antiferromagnetic quantum-critical point in itinerant fermions as well as to incommensurate Ising model in any of is physical applications.

We are interested in the fluctuations in which 
$(\phi_x, \phi_y)({\bf r}, \tau)$ is space and time dependent and small. Consistent with the XY model assumption for the model, the eigenvalues of the effective Hamiltonian are 
invariant to values of $(\phi_x, \phi_y)$ as long as they lie on a circle.  (The case of coupling to fermions including the fourfold lattice anisotropies is given
in Ref. \cite{ASV2010}.) Consider radial variations $\delta \phi (r)$ and the orthogonal variations,
\be
\left(\begin{array}{ccc}\delta \phi_x(r) \\ \delta \phi_y(r)\end{array}\right) = \left(\begin{array}{ccc} - {\hat{y}} \\ {\hat{x}}\end{array}\right) \delta \theta(r).
\ee
We expand for slow and small spatial variations in
$\delta \phi(r)$. Schematically,
$$\epsilon(k + i \delta \phi(r)) \approx \epsilon(k) + i(d\epsilon(k)/d \phi) \delta \phi(r) = \epsilon(k) + i(d\epsilon(k)/dk) \delta \phi(r).$$
The slow variations in $\delta \phi({\bf r})$ lead to two different forms of coupling $i = r$ for radial variations and $i = t$ for tangential variations,
\be
\delta H &=& \sum_{{\bf k,q}}\sum_{i = r,t} \delta \theta({\bf q}, i)_r \gamma^i({\bf k, k+q}) c_{{\bf k+q}}^+ c_{\bf k} + H.C., \\ 
\gamma^r({\bf  k, k+q}) &=& i~ (\epsilon_{{\bf k+q}} - \epsilon_{{\bf k}}) \approx i~q v({\bf k}_F), \\
\gamma^t({\bf  k, k+q}) &=& \frac{1}{2} i \Big( - \frac{d\epsilon}{dk'_x}{\hat{k}_{y}}+ \frac{d\epsilon}{dk'_y} {\hat{k_x}}\Big) + ~ interchange ~(k, k').
\ee
We note that for a circular Fermi surface, we may write
\be
\label{gk}
\gamma^t({\bf k, k'}) = i \gamma ({\bf k} \times {\bf k}').
\ee
The longitudinal or radial coupling  of the fluctuations  $\delta \theta_{\ell} ({\bf q}) = {\bf q} \theta ({\bf q})$ couple to the Fermions current ${\bf j}({\bf k_F})$. This is indeed how a collective "spin-wave" current $\nabla \theta({\bf r})$ should couple to fermions. But, it is important to note that the fluctuations of $\theta({\bf r}, t)$ in space and time are not the critical fluctuations of the loop-current order. They are however finite and have been used however to generate $S_{c-f}$ by integrating over the fermion currents and give the result shown in Eq. (\ref{scfqw}). The transverse couplings in a similar calculation also give a contribution to $S_{c-f}$ of precisely the same functional form.

For transverse or tangental coupling, the fermion operator $c_{\bf k} i ({\bf k} \times {\bf k}') c_{\bf k}'$ is the fermion angular momentum operator. Therefore $\delta\theta_{r} ({\bf q}) = -i \frac{\partial \theta}{\partial {\hat{\bf z}}} {\bf q}  = {\bf L}_z {\bf q}$, the angular momentum operator is the generator of rotations of $\theta({\bf r}, t)$. As already mentioned earlier the correlations of $L_z({\bf r}, t)$ are also the critical mode of the quantum XY model. Therefore they are to be used for calculating the single-particle self-energy, the resistivity and other transport properties. Remarkably, even though the single-particle relaxation rate in the normal state in such calculations is only weakly angular dependent, they lead to attraction in the $d$-wave single-particle channel since they lead to preferential scattering of fermions through an angle $\pm \pi/2$ \cite{ASV2010} (see also \cite{Bok_ScienceADV}).

It is worth noting also that the quantum Monte Carlo calculations \cite{ZhuHouV2016} show that the coupling of $L_z(r,t)$ (or of $\cos \theta(r,t)$) to fermions to generate an effective term in the action for the collective variables is irrelevant. Inclusion of only such a term keeps the fluctuations of the Lorentz-invariant form, Eq. (\ref{D0}). This leaves only the longitudinal coupling to generate $S_{c-f}$, which is what has been done.

\section{Single-particle self-energy and vertex correction to it} 

In the MFL phenomenology, a momentum independent fluctuation spectrum and a momentum independent coupling of the fluctuations to the fermions were assumed. Then a  momentum independent single-particle self-energy is obtained. It is then easy to show that the vertex correction to the self-energy goes to zero at low energies and near the Fermi momentum. Because of the $\omega/T$ dependence of the fluctuations the imaginary part of the retarded self-energy is proportional to $|\omega|$. Every transport property calculated in this paper then follows. 

 In the microscopic theory, the frequency and the temperature dependence of the fluctuations is the same; they are also of product form in their $\omega/T$ dependence and in their momentum dependence. But the fluctuations have a singular dependence at low momenta and the coupling function to fermions is also momentum dependent. We now give results for the self-energy in the microscopic theory and show that to leading order in $(\omega, T)$, the results remain the same. It should be noted that while precise results can be given for a circular Fermi surface with the assumption of momentum-independent fluctuations and vertex to fermions, it is not so for a realistic Fermi surface and with the results of the microscopic theory for the fluctuations and the coupling to fermions.

Using Eq.~(\ref{Dqw}) for the propagator of fluctuations and Eq.~(\ref{gk}) for the coupling function of the fluctuations to the fermions, the single-particle self-energy given by Fig. \ref{Fig:SE}(c) without the triangular vertex in it (to which we will come back), i.e. just the second order in the coupling 
without vertex corrections to the self-energy, is given by
\be
{\rm Im} \Sigma_R({\bf k}, \varepsilon) &=& \frac{1}{2V} \cosh\frac{\varepsilon}{2T} 
\sum_{\bf k'} \int_{-\infty}^{\infty} d \varepsilon' |g({\bf k},{\bf k}')|^2 
{\rm Im} D_R({\bf k-k}', \varepsilon - \varepsilon') A({\bf k}', \varepsilon') \nonumber\\
&\quad& \times  {\rm cosech}\frac{\varepsilon -\varepsilon'}{2T}{\rm sech}\frac{\varepsilon'}{2T}.
\label{se}
\ee
We note that for $A({\bf k}, \varepsilon) \equiv (-1/\pi){\rm Im} G_R({\bf k}, \varepsilon)=a(\varepsilon) \delta(\varepsilon- \epsilon_{\bf k}) = \delta(a^{-1} \varepsilon - \tilde{\epsilon}_{{\bf k}})$ with a self-consistent $G_R$, $\tilde{\epsilon}_{{\bf k}}$ is the energy renormalized by the quasiparticle weight, measured from the chemical potential.  Let us further define $\varepsilon = t T$. Then using that $\delta(a^{-1} t' T - \tilde{\epsilon}_{{\bf k}'})$ and that $a^{-1} t' T$ is negligible at low temperatures, we get after the energy integral that
\be
\label{Sig1}
{\rm Im} \Sigma_R({\bf k}, \varepsilon) = - \overline{g}_0^2N(0) I_0({\bf k}) \times
\left\{
\begin{array}{ll}
\displaystyle{\frac{1}{2} (\pi^2 T^2 + \varepsilon^2)} \xi_{\tau} & \qquad \mbox{for $2T\xi_\tau \ll 1$}
\\
&
\\
\displaystyle{\varepsilon \coth \frac{\varepsilon}{2T}} & \qquad \mbox{for $2T\xi_\tau \gg 1$}
\end{array}
\right. ,\ee
where $\overline{g}_0^2=D_0\xi_r^{-2} |g({\bf k},{\bf k}-{\bf q})|^2/ 2|{\bf \hat{k}} \times {\bf q}|^2$.

The momentum dependence of the self-energy given by
\be
\label{Ik}
I_0({\bf k}) = \frac{2}{N(0)V} \sum_{{\bf q}} \frac{\theta(q_c-q)}{q^2 + \xi_r^{-2}} 
 |{\bf \hat{k}} \times {\bf q}|^2\delta(\tilde{\epsilon}_{{\bf k}-{\bf q}}), 
\ee
where $q_c$ is a cutoff wave number, depends on the Fermi surface. 

Let us take a model Fermi surface near criticality devised to fit that determined experimentally by ARPES \cite{AbrahamsV_Hall_2001, AbrahamsV_Hall_2003}}. This form had been devised to calculate aspects of the measured Hall effect and impurity resistivity caused by scattering through small angles by impurities in between the Cu-O planes. It is enough to specify the angular dependence of the Fermi velocity $v(\theta_v)$ with respect to the  axes of the Brillouin zone with the $(\pi,\pi)$ point taken as the origin,
\be
{\bf v}_F(\theta_v) &=& v_x(\theta_v) {\hat{x}} + v_y(\theta_v) {\hat{y}}\\
v_x(\theta_v) &=&v_0(\sin \theta_v  + \rho \sin 3\theta_v )\\
v_y(\theta_v) &=& v_0(\cos \theta_v  - \rho \cos 3\theta_v ). 
\ee
$\rho = 0$ gives a circular Fermi surface. The actual value of $\rho$ may be determined from experiments. One needs only to specify the Fermi vector at $\theta_v  = 0$ to specify  the entire Fermi surface. 
We calculate expanding the energy 
$\tilde{\epsilon}_{{\bf k}-{\bf q}} \simeq \tilde{\epsilon}_{\bf k} - {\bf v}({\bf k}) \cdot {\bf q}$ linearly in momentum ${\bf q}$ across the Fermi surface ${\bf k}_F$ and imposing an upper cutoff $q_c$ through $W_c = q_c |v(\theta_v)|$. Defining the angle $\theta_{v_{k,k}}$ as the angle between ${\bf v}({\bf k})$ and ${\bf k}$, we find
\be
\label{Ik2}
I({\bf k}) &\simeq& \frac{1}{4\pi v_{\bf k}^2}
\left[\left( \frac{2W_c(\theta_v)}{\pi} - \sqrt{\tilde{\epsilon}_{\bf k}^2 + v_{\bf k}^2 \xi_r^{-2}} \right)
\cos^2 \theta_{v_{k,k}} 
+ \frac{\tilde{\epsilon}_{\bf k}^2}{\sqrt{\tilde{\epsilon}_{\bf k}^2 + v_{\bf k}^2 \xi_r^{-2}}}
\sin^2 \theta_{v_{k,k}}\right] .
\ee
Here $W_c(\theta_v)$ is the upper cutoff in the energy of order the bandwidth, which in general depends on the angle on the Fermi surface. For a circular Fermi surface, $ \theta_{v_{k,k}} =0$  and $k/ v_{{\bf k}} = m$. 
For $\xi_r \to \infty$, i.e. precisely at the critical point, $\sqrt{\tilde{\epsilon}_{\bf k}^2 + v_{\bf k}^2 \xi_r^{-2}}$ is  $\tilde{\epsilon}_{\bf k}$ for $k > k_F$ but $-\tilde{\epsilon}_{\bf k}$ for $k < k_F$, 
so that $I({\bf k})$ has a cusp at $k=k_F$. This is so for a general Fermi surface. But the leading term is given by the bandwidth divided by the square of the Fermi velocity at ${\bf k}$. Moreover, {\it except exponentially close to criticality}, $\xi_r^{-1}$ is of order $k_F$, so that the singularity is unimportant and in any case, and hard to observe in experiments. Expanding about the Fermi momentum, the leading momentum dependence is a constant  and the correction is of $O({\bf (k-k_F)/k_F})$.

\subsection{Vertex correction for self-energy}

\begin{figure}
\begin{center}
\includegraphics[width= 0.8\columnwidth]{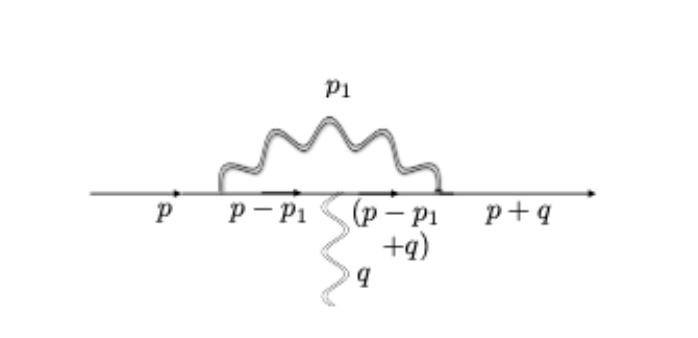}
\end{center}
\caption{Leading vertex correction}
\label{Fig:vertex}
\end{figure}

We estimate the leading vertex correction to the self-energy given by the diagram (\ref{Fig:vertex}).  A result much stronger than the Migdal vertex correction in the electron-phonon interactions is obtained, that the vertex correction for energy and momentum $p$ near the Fermi energy goes to $0$ linearly, because of the separability of the fluctuation propagator as  a function of energy and of momenta. As in Migdal's argument, any higher-order vertex correction is  expected to be even more unimportant. 

The vertex correction of Fig. (\ref{Fig:vertex}) is given by
\be
\label{vert1}
\gamma^{(2)}(p,q) = \sum_{p_1,q_1}  g^2\sum_{p_1}\frac{|{\bf p \times p}_1|^2}{|{\bf p}_1|^2+ \xi_r^{-2}} D(p_1) G_0(p-p_1) G_0(p-p_1+ q).
\ee
$(p,q, p_1,q_1)$ etc. are, for example $p= ({\bf p},p+i sgn(p))$, and $D(p_1)$ denotes only the frequency-dependent part of the fluctuation spectra.

For the calculation of the self-energy (and transport), one needs results for a general $(q_0,{\bf q})$, with $(p_0, {\bf p-p}_F)$ near zero. We evaluate Eq. (\ref{vert1}) by noting that given the separation of energy and momentum in the product of the bare vertex and the fluctuation propagator $D$, it is simpler to first do the integration of the product of the fermion Green's function over momentum ${\bf p}_1$.  To do the momentum integrals,  change variables to the energies $\epsilon(|{\bf p-p}_1|)$ and the angle between $({\bf p-p}_1)$ and ${\bf p}$ and similarly for $\epsilon(|{\bf p-p}_1+{\bf q}|)$. Now we may safely take the integrals over the energies on the real axis from $-\infty$ to $\infty$ and do a contour integral. The unperturbed Green's functions have poles such that the contours must be closed in opposite directions to get nonzero results. This procedure gives
\be
\label{ver2}
\gamma^{(2)}(p,q) = \frac{i m}{2\pi} \sum_{p_1} g^2\frac{ |{\bf p \times p_1}|^2}{|{\bf p}_1|^2+ \xi_r^{-2}} D(p_1) \frac{\theta(p_0-p_{10}) - \theta (p_0-p_{10} +q_0)}{q_0 -{\bf q}.{\bf v}_{\bf p-p_1} + i sgn(q_0)}. 
\ee
The integrand over the frequency $p_1$  is nonzero only between $p_0$ and $(p_0+q_0)$. This shows that the vertex correction is linear in $p_0$ and $q_0$, which proves its unimportance for leading frequency and temperature dependence in self-energy, provided the remaining integral has no divergence. Integrating over the frequency part of $p_1$ (for $p_0, q_0$ both $>0$, and similar results elsewhere),
\be
\gamma^{(2)}(p,q) \propto &~ g^2 \Big( i \big(p_0 \ln \frac{|p_0 + q_0|}{p_0} + q_0 (\ln\frac{|p_0+q_0|}{\omega_c} -1)\big) + q_0\Big) \\ \nonumber
&\times \sum_{\bf p_1} \frac{ |{\bf p \times p_1}|^2}{|{\bf p}_1|^2+ \xi_r^{-2}} \frac{1}{{\bf v}_{\bf p-p_1} + i sgn(q_0)}.
\ee
The momentum integral at its maximum, i.e. at $\xi_r^{-1} \to \infty$, and for ${\bf p-p}_1 = {\bf p}_F$, gives $\propto 1/q_0$ for $q_0 \to 0$,  which is of no significance since $q_0$ is summed over the whole range of energy. For $q_0^2 > v_f^2 |{\bf q}|^2$, it gives 
$\propto \frac{1}{v_f^2 |{\bf q}|^2} (q_0- sgn(q_0)\sqrt{q_0^2-v_f^2 |{\bf q}|^2})$, which  multiplied by $p_0$  does not correct the leading frequency or temperature dependence of the self-energy, especially as $(q_0, {\bf q})$ are integrated over.

We have therefore proved that in the present problem, there is no vertex correction to leading order in $(\omega,T, {\bf p-p}_F)$ to the self-energy, which is the region in which the transport properties are derived.

\section{Evaluations of the memory matrix elements} 

In the main text we have shown that the memory matrix is a sum of three contributions, from the self-energy, the Maki-Thompson type diagram and the Aslamazov-Larkin type diagram. Their matrix elements are given by
\begin{align}
[\hat{M}_{\rm SE}''(t, t')]_{LL'} 
=& - \frac{T}{N(0)} \int \frac{d {\bf p}\,d {\bf p}'}{(2\pi)^{2d}} 
\psi_{L}({\bf p})  \psi_{L'}({\bf p}) \delta({\tilde{\epsilon}}_{\bf p}) 
\delta({\tilde{\epsilon}}_{{\bf p}'}) 
\nonumber\\
& \times
\int_{-\infty}^{\infty} dx
|g({\bf p}, {\bf p}')|^2 
{\rm Im} D_R({\bf p}- {\bf p}', Tx) 
\nonumber \\
& \times
\delta(t-t')  \cosh \frac{t}{2} \, {\rm sech} \frac{t-x}{2} \, {\rm cosech} \frac{x}{2} ,
\label{memory_SE_FSH}
\\
[\hat{M}_{\rm MT}''(t, t')]_{LL'} 
=&\frac{T}{N(0)} \int \frac{d {\bf p}\,d {\bf p}'}{(2\pi)^{2d}} 
\psi_{L}({\bf p}) \delta({\tilde{\epsilon}}_{\bf p})  \delta({\tilde{\epsilon}}_{{\bf p}'}) \psi_{L'}({\bf p}') 
\nonumber \\
& \times
|g({\bf p}, {\bf p}')|^2 {\rm Im} D_R({\bf p}- {\bf p}', T t-T t') 
\nonumber \\
&\times
\,{\rm cosech} \frac{t-t'}{2} ,
\label{memory_MT_FSH}
\\
[\hat{M}_{\rm AL}''(t, t')]_{LL'} 
=& \frac{T}{2 N(0)} \int \frac{d {\bf p}\,d {\bf p}'}{(2\pi)^{2d}}  
\psi_{L}({\bf p}) \delta({\tilde{\epsilon}}_{\bf p})  \delta({\tilde{\epsilon}}_{{\bf p}'}) 
\nonumber \\
& \times
\int \frac{d {\bf k}d {\bf k}'}{(2\pi)^{d}}  
\Delta^d ({\bf p}-{\bf p}'+{\bf k}-{\bf k}') 
\delta({\tilde{\epsilon}}_{\bf k})  \delta({\tilde{\epsilon}}_{{\bf k}'}) \psi_{L'}({\bf k}')
\nonumber \\
& \times
\left[\int \frac{d {\bf k}_1d {\bf k}_1'}{(2\pi)^{d}}  
\Delta^d ({\bf p}-{\bf p}'+{\bf k}_1-{\bf k}_1') 
\delta({\tilde{\epsilon}}_{{\bf k}_1})  \delta({\tilde{\epsilon}}_{{\bf k}_1'}) \right]^{-1}
\nonumber \\
& \times
\int_{-\infty}^{\infty} dx 
|g({\bf p}, {\bf p}')|^2 {\rm Im} D_R({\bf p}-{\bf p}',T x) 
\nonumber \\
& \times \frac{1}{x}\, {\rm sech} \frac{t-x}{2}\left({\rm sech} \frac{t'+x}{2}
+\, {\rm sech} \frac{t'-x}{2}
\right) .
\label{memory_AL_FSH}
\end{align}
Here $\Delta^d ({\bf p}-{\bf p}'-{\bf q})  \equiv \sum_{\bf G} \delta^d ({\bf p}-{\bf p}'-{\bf q}-{\bf G})$ 
is a $d$-dimensional delta function extended to the lattice with ${\bf G}$ being the reciprocal lattice vectors 
including a zero vector, and then normal and Umklapp scatterings are described by ${\bf G} = {\bf 0}$ and ${\bf G} \neq {\bf 0}$, respectively.

The momentum independence of Eq.~(\ref{assumption_fluctuation}) leads to the fact that $\hat{M}''_{\rm SE}(t,x)$ is proportional to the unit matrix $\hat{1}$ whose matrix element is $\delta_{L,L'}$ and 
that $\hat{M}''_{\rm MT}(t,x)$ vanishes. Then $\hat{M}''_0(t,x) =\hat{M}''_{\rm SE}(t,x) +\hat{M}''_{\rm AL}(t,x)$, 
and Eq.~(\ref{memory_SE_FSH}) with Eq.~(\ref{assumption_fluctuation}) leads to
\begin{align}
\hat{M}''_{\rm SE}(t,x) &= 2 \bar{g}^2 N(0)T f(t) \delta(t-x) \hat{1}, 
\label{memory_SE_LCO}
\end{align}
where
\begin{align}
f(t) &= t \coth (t/2) .
\end{align}
The AL vertex corrections can be obtained by substituting  Eq.~(\ref{assumption_fluctuation}) into Eq.~(\ref{memory_AL_FSH}), as follows:
\begin{align}
\hat{M}''_{\rm AL}(t,x)&=
- \bar{g}^2 N(0)T [ F(t,x) + F(t,-x) ] \hat{B} ,
\label{memory_AT_LCO}
\end{align}
where
\be
F(t,x) &=& \frac{1}{4} 
\int_{-\infty}^{\infty}\frac{dy}{f(y)} \,{\rm sech} \frac{t-y}{2}  \,{\rm sech} \frac{y-x}{2},
\label{funcF}
\ee
and the matrix elements of $\hat{B}$ are given by
\be
B_{LL'} &=& \frac{2}{N(0)^2} 
\int \frac{d {\bf p}d {\bf p}'}{(2\pi)^{2d}}  
\psi_{L}({\bf p}) \delta({\tilde{\epsilon}}_{\bf p})  \delta({\tilde{\epsilon}}_{{\bf p}'}) 
\nonumber \\
&\quad& \times
\int \frac{d {\bf k}d {\bf k}'}{(2\pi)^{d}}  
\Delta^d ({\bf p}-{\bf p}'+{\bf k}-{\bf k}') 
\delta({\tilde{\epsilon}}_{\bf k})  \delta({\tilde{\epsilon}}_{{\bf k}'}) \psi_{L'}({\bf k}')
\nonumber \\
&\quad& \times
\left[\int \frac{d {\bf k}_1d {\bf k}_1'}{(2\pi)^{d}}  
\Delta^d ({\bf p}-{\bf p}'+{\bf k}_1-{\bf k}_1') 
\delta({\tilde{\epsilon}}_{{\bf k}_1})  \delta({\tilde{\epsilon}}_{{\bf k}_1'}) \right]^{-1}.
\nonumber\\
\label{vertex matrix}
\ee
Then the effect of the vertex corrections is described by the matrix $\hat{B}$. Gauge invariance relates the vertex corrections to the self-energy as Eq.~(\ref{Ward--Takahashi}). This relation corresponds to the identity
\begin{align}
\int_{-\infty}^{\infty} \frac{F(t,x)}{\cosh(x/2)} dx = \frac{f(t)}{\cosh(t/2)} .
\label{gauge inv}
\end{align}
Now, following the exact solution of the Boltzmann equation by Jensen, Smith, and Wilkins~\cite{Jensen-S-W1969}, we consider the eigenvalue equation 
\be
f(t) \varphi_{n,\pm}(t) = \lambda_{n,\pm} \int_{-\infty}^{\infty} dx F(t,x) \varphi_{n,\pm}(x) ,
\label{eigenvalue}
\ee
where $\varphi_{n,+}(t)$ and $\varphi_{n,-}(t)$ are even and odd functions of $t$, respectively, $\varphi_{n,\pm}(t) = \pm \varphi_{n,\pm}(\pm t) $, and they are normalized according to 
\begin{align}
\int_{-\infty}^{\infty} dt f(t) \varphi_{n,s}(t) \varphi_{n',s'}(t)  = \delta_{n,n'} \delta_{s,s'}.
\end{align}
These eigenfunctions and eigenvalues are given by
\be
\varphi_{n,+} (t) &=& \frac{1}{\pi} {\cal P}_{2n-2}(t/\pi)\, {\rm sech} \frac{t}{2}, 
\quad \lambda_{n,+} = (2n-1)^2, \qquad\\
\varphi_{n,-} (t) &=& \frac{1}{\pi} {\cal P}_{2n-1}(t/\pi)\, {\rm sech} \frac{t}{2}, 
\quad \lambda_{n,-} = (2n)^2,
\ee
where ${\cal P}_n(x)$ are orthogonal polynomials in the interval $(-\infty, \infty)$ with a weight function $2x\,{\rm cosech} (\pi x)$, for example, ${\cal P}_0(x) = 1$, ${\cal P}_1(x) = \sqrt{2} x$, and ${\cal P}_2(x) = 1-2x^2$. 
Note that from Eq.~(\ref{funcF}) they also satisfy the following auxiliary eigenequation:
\be
f(t) \varphi_{n,\pm}(t) = \frac{\sqrt{\lambda_{n,\pm}}}{2} \int_{-\infty}^{\infty} dx 
\,{\rm sech} \frac{t-x}{2}  \varphi_{n,\pm}(x) .\quad
\ee 
All the eigenvalues are grater than or equal to $1$ and the eigenequation for $\varphi_{1,+}(t) = (1/\pi) \, {\rm sech} (t/2)$ with the smallest eigenvalue $\lambda_{1,+}=1$ corresponds to the gauge invariance, Eq.~(\ref{gauge inv}).  

Using the eigenvalue $\lambda_{n,\pm} \geq 1$ and the eigenfunctions $\varphi_{n,\pm}(t)$, we can write Eqs.~(\ref{memory_SE_LCO}) and (\ref{memory_AT_LCO}) as
\begin{align}
\hat{M}''_{\rm SE}(t,x) &= 2 \bar{g}^2 N(0)T f(t)f(x)
\sum_{n=1}^{\infty} \sum_{s=\pm} \varphi_{n,s}(t) \varphi_{n,s}(x)\, \hat{1},
\\
\hat{M}''_{\rm AL}(t,x) &=
- 2 \bar{g}^2 N(0)T f(t)f(x) 
\sum_{n=1}^{\infty} \frac{\varphi_{n,+}(t) \varphi_{n,+}(x)}{\lambda_{n,+}} \, \hat{B} .
\end{align}
Therefore we obtain the imaginary part of the memory matrix, $\hat{M}_0''(t,x) = \hat{M}''_{\rm SE}(t,x)  + \hat{M}''_{\rm AT}(t,x)$, for the MFL 
\begin{widetext}
\be
\hat{M}_0''(t,x) &=&
2 \bar{g}^2 N(0)T f(t)f(x) 
\sum_{n=1}^{\infty} 
\left(
\frac{\varphi_{n,+}(t) \varphi_{n,+}(x)}{\lambda_{n,+}} \,( \lambda_{n,+} \hat{1} - \hat{B} )
+ \varphi_{n,-}(t) \varphi_{n,-}(x)\, \hat{1}
\right).
\nonumber \\
\label{memory LCO}
\ee
The inverse matrix can be easily obtained as
\be
\hat{M}_0''(t,x)^{-1} &=&
\frac{1}{2 \bar{g}^2 N(0)T} 
\sum_{n=1}^{\infty}
\left(
\varphi_{n,+}(t) \varphi_{n,+}(x) \lambda_{n,+} \,( \lambda_{n,+} \hat{1} - \hat{B} )^{-1} 
+ \varphi_{n,-}(t) \varphi_{n,-}(x)\, \hat{1}
\right).
\nonumber \\
\label{inverse memory}
\ee
\end{widetext}
Since $\hat{M}''(t,x)^{-1} \propto 1/T$, we see that the critical fluctuations of the QXY-F model give rise to the Planckian dissipation where the transport relaxation time is proportional to $\hbar/k_{\rm B} T$ 
(we have reinserted the Planck constant $\hbar$ and the Boltzmann constant $k_{\rm B}$).
 
Substituting Eq.~(\ref{inverse memory}) into Eqs.~(\ref{sigma_FSH}) and (\ref{kappa_memory}), we obtain.
\be
\sigma &=& 
\frac{e^2\langle {\tilde v}_x^2 \rangle}{\bar{g}^2T} 
\sum_{n=1}^{\infty} \left( \int_{-\infty}^{\infty}  \frac{\varphi_{n,+}(t) }{2 \cosh (t/2)} dt \right)^2
\left[ \frac{\lambda_{n,+}}{\lambda_{n,+} \hat{1} - \hat{B}} \right]_{11} ,
\label{sigma_LCO}
\\
\kappa &=& 
\frac{\langle {\tilde v}_x^2 \rangle}{\bar{g}^2} 
\sum_{n=1}^{\infty}
\left( \int_{-\infty}^{\infty}  \frac{t \varphi_{n,-}(t) }{2 \cosh (t/2)} dt \right)^2 .
\label{}
\ee
Note that the effect of the vertex corrections $\hat{B}$ vanishes for the thermal conductivity $\kappa$ 
owing to the symmetry reasons mentioned at the end of Sec. III,
while the electrical conductivity has the vertex corrections to the self-energy contribution, 
\be
\sigma_{\rm SE} =
\frac{e^2\langle {\tilde v}_x^2 \rangle}{\bar{g}^2T} 
\sum_{n=1}^{\infty} \left( \int_{-\infty}^{\infty}  \frac{\varphi_{n,+}(t) }{2 \cosh (t/2)} dt \right)^2.
\label{sigma_SE}
\ee 
Using the completeness 
\be
f(t) \sum_{n=1}^{\infty} \sum_{s=\pm} \varphi_{n,s}(t) \varphi_{n,s}(x) = \delta(t-x),
\ee
$\sigma_{\rm SE}$ and the thermal conductivity can be calculated as
\be
\sigma_{\rm SE} &=&
\frac{e^2\langle {\tilde v}_x^2 \rangle}{\bar{g}^2T} 
\int_{-\infty}^{\infty}  \frac{\tanh (t/2)}{4 t \cosh^2 (t/2) f(t) } dt
= \frac{7 \zeta(3)}{2 \pi^2} \frac{e^2\langle {\tilde v}_x^2 \rangle}{\bar{g}^2T} ,
\label{result_sigmaSE}
\\
\kappa &=& 
\frac{\langle {\tilde v}_x^2 \rangle}{\bar{g}^2} 
\int_{-\infty}^{\infty}  \frac{t \tanh (t/2) }{4 \cosh^2 (t/2)} dt 
= \frac{\langle {\tilde v}_x^2 \rangle}{\bar{g}^2} .
\label{result_kappa}
\ee
Because of the Planckian dissipation, the thermal conductivity is a constant independent of temperature. 
The Lorentz number $L$ can then be described by
\be
L \equiv \frac{\kappa}{\sigma T} = \frac{\kappa}{\sigma_{\rm SE}T} 
\frac{\rho}{\rho_{\rm SE}}
= \frac{6}{7\zeta(3)} \frac{\rho}{\rho_{\rm SE}} L_0 ,
\ee
where $\rho = 1/\sigma$, $\rho_{\rm SE}=1/\sigma_{\rm SE}$, and $L_0 = (\pi^2/3) (k_{\rm B}/e)^2$ is the Lorentz number of normal metals. Since $\frac{6}{7\zeta(3)}=0.713$, the value of $L/L_0$ is about 70\% of the value of $\rho/\rho_{\rm SE}$.

By using Eq.~(\ref{sigma_SE}), we can write Eq.~(\ref{sigma_LCO}) as
\be
\frac{\sigma}{\sigma_{\rm SE}} &=& 1+
\frac{2 \pi^2} {7 \zeta(3)}
\sum_{n=1}^{\infty} \left( \int_{-\infty}^{\infty}  \frac{\varphi_{n,+}(t) }{2 \cosh (t/2)} dt \right)^2
\nonumber\\&\quad&\times
\left[  \frac{\lambda_{n,+} }{(\lambda_{n,+} \!-1) \hat{1} + \hat{C}} - \hat{1} \right]_{11} ,
\label{resistivity_ratio}
\ee
where 
\be
\label{hatC}
\hat{C} \equiv \hat{1} - \hat{B}.
\ee
If we keep only the $n=1$ term in the sum over $n$ in Eq.~(\ref{resistivity_ratio}), 
we obtain the inequality
\be
\frac{\sigma}{\sigma_{\rm SE}}
>  1- \frac{8} {7 \zeta(3)} + \frac{8} {7 \zeta(3)} [\hat{C}^{-1}]_{11}.
\ee
By replacing all eigenvalues $\lambda_{n,+}$ with the minimum eigenvalue $\lambda_{1,+} = 1$, 
on the other hand, we obtain the inequality
\be
\frac{\sigma}{\sigma_{\rm SE}}
< [\hat{C}^{-1}]_{11}.
\ee
Therefore the lower and upper bounds of $\rho/\rho_{\rm SE}$ are given by
\be
\left( \frac{\rho}{\rho_{\rm SE}} \right)_{\rm L.B.}
&=& \frac{1}{[\hat{C}^{-1}]_{11}},
\label{Umklapp_LB}
\\
\left( \frac{\rho}{\rho_{\rm SE}} \right)_{\rm U.B.}
&=& \frac{7 \zeta(3)/8}{[\hat{C}^{-1}]_{11}+7 \zeta(3)/8-1}.
\label{Umklapp_UB}
\ee
Since $7 \zeta(3)/8=1.0518$, however, these bounds are almost equal, so that we can take the lower bound for evaluating the electrical resistivity. The results for its explicit evaluation for the case of a circular Fermi surface in a square lattice are given 
in Sec. IV C. 

As described in Sec. II C, 
if there is a conserved quantity that is odd in time reversal and in inversion, the imaginary part of the memory matrix has a zero eigenvalue as Eq.~(\ref{zeroeigenvalue}). By substituting Eq.~(\ref{memory LCO}) for Eq.~(\ref{zeroeigenvalue}) while noting that $\varphi_{1,+}(t) \propto {\rm sech}(t/2)$, this corresponds to the matrix ${\hat C}$ having a zero eigenvalue,
\be
{\hat C}|\theta \rangle = 0.
\ee
Thus, ${\hat C}$ is a matrix describing conservation laws. For the lower bound of $\rho/\rho_{\rm SE}$, we can easily see the relationship between the conservation laws and particle-hole symmetry, since the replacement of all $\lambda_{n,+}$ by 1 in Eq.~(\ref{memory LCO}) leads to
\be
\hat{M}_0''(t,x) &\approx&
2 \bar{g}^2 N(0)T f(t) 
\left\{ \left[\delta(t-x)+\delta(t+x) \right] {\hat C}\right.
\left. + \left[\delta(t-x)-\delta(t+x)\right] {\hat 1}
\right\}.
\ee
On the other hand, the upper bound, Eq.~(\ref{Umklapp_UB}), shows that $\rho/\rho_{\rm SE}$ vanishes in correspondence with the zero eigenvalues of ${\hat C}$.

\section{Umklapp factor for a circular Fermi surface} 

\begin{figure}[t]
\begin{center}
\includegraphics[width=0.8\columnwidth]{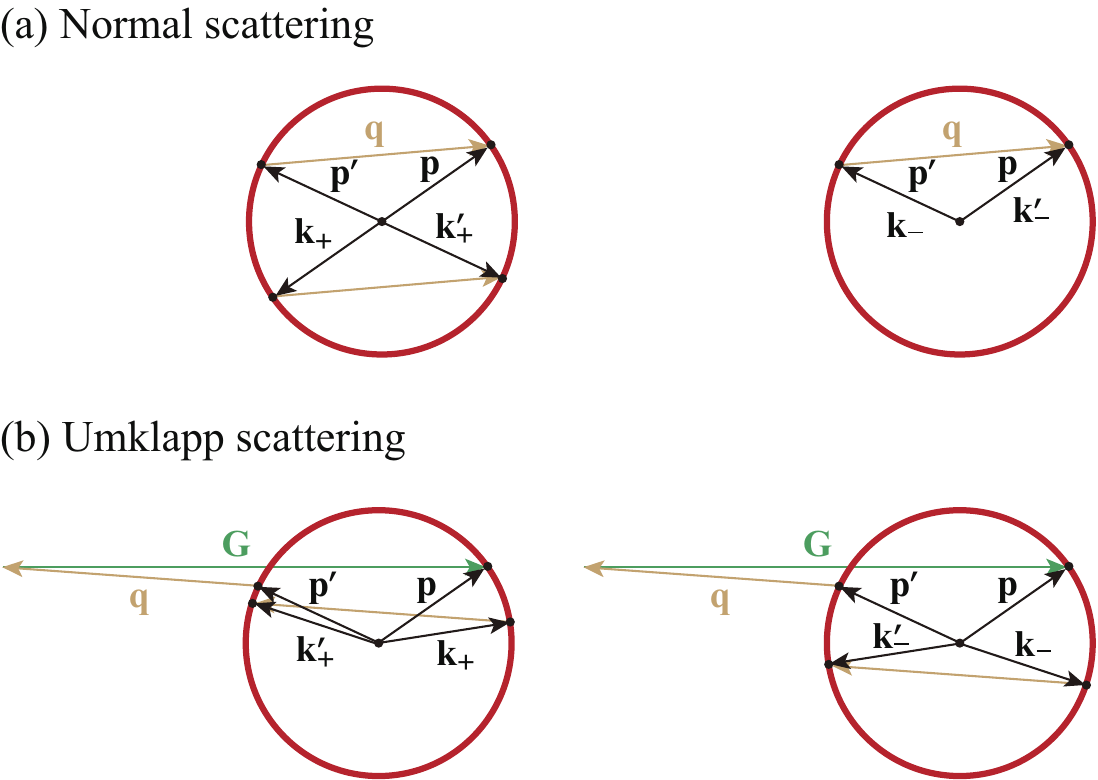}
\end{center}
\caption{Normal and Umklapp scattering processes for a two-dimensional circular Fermi surface.  
For the given ${\bf p}$, ${\bf p}'$, and ${\bf G}$, there are only two processes that satisfy 
${\bf p} - {\bf p}' + {\bf k} - {\bf k}' = {\bf G}$, indicated by $\pm$.
These processes are related by ${\bf k}_\pm' = - {\bf k}_\mp$.
 (a) For normal scattering (${\bf G} = {\bf 0}$), ${\bf k}_+ = - {\bf p}$, ${\bf k}_+' = - {\bf p}'$, 
${\bf k}_- = {\bf p}'$ and ${\bf k}_-' = {\bf p}$, 
so that $\psi_{L'}({\bf p}) - \psi_{L'}({\bf p}')  + \psi_{L'}({\bf k}_\pm) - \psi_{L'}({\bf k}_\pm')$ 
in Eq.~(\ref{element_C}) is identically zero. 
(b) For Umklapp scattering (${\bf G} \neq {\bf 0}$), 
$\psi_{L'}({\bf k}_+) - \psi_{L'}({\bf k}_+') = \psi_{L'}({\bf k}_-) - \psi_{L'}({\bf k}_-')$
but  $\psi_{L'}({\bf p}) - \psi_{L'}({\bf p}')  + \psi_{L'}({\bf k}_\pm) - \psi_{L'}({\bf k}_\pm')$ becomes nonzero.}
\label{Fig:Scatt}
\end{figure}

From Eq.~(\ref{vertex matrix}) with use of $\psi_{L}(-{\bf p}) = - \psi_{L}({\bf p})$, 
the matrix elements of $\hat{C} = \hat{1} - \hat{B}$ are obtained as 
\be
C_{LL'} &=& \frac{1}{N(0)^2} 
\int \frac{d {\bf p}d {\bf p}'}{(2\pi)^{2d}}  
\psi_{L}({\bf p}) \delta({\tilde{\epsilon}}_{\bf p})  \delta({\tilde{\epsilon}}_{{\bf p}'}) 
\nonumber \\
&\quad& \times
\int \frac{d {\bf k}d {\bf k}'}{(2\pi)^{d}}  
\Delta^d ({\bf p}-{\bf p}'+{\bf k}-{\bf k}')  
\delta({\tilde{\epsilon}}_{\bf k})  \delta({\tilde{\epsilon}}_{{\bf k}'}) 
\nonumber \\
&\quad& \times
\left[\psi_{L'}({\bf p}) - \psi_{L'}({\bf p}')  + \psi_{L'}({\bf k}) - \psi_{L'}({\bf k}') \right]
\nonumber \\
&\quad& \times
\left[\int \frac{d {\bf k}_1d {\bf k}_1'}{(2\pi)^{d}}  
\Delta^d ({\bf p}-{\bf p}'+{\bf k}_1-{\bf k}_1')  
\delta({\tilde{\epsilon}}_{{\bf k}_1})  \delta({\tilde{\epsilon}}_{{\bf k}_1'}) \right]^{-1}.
\nonumber\\
\label{element_C}
\ee
Since $\Delta^d ({\bf p}-{\bf p}'+{\bf k}-{\bf k}')  
= \sum_{\bf G} \delta^d ({\bf p}-{\bf p}'+{\bf k}-{\bf k}'-{\bf G})$, 
the matrix can be separated as $\hat{C} = \hat{C}_N + \hat{C}_U$, 
where $\hat{C}_N$ and $\hat{C}_U$ are the contributions from normal (${\bf G} = {\bf 0}$) 
and Umklapp (${\bf G} \neq {\bf 0}$) scatterings, respectively. 
$\hat{C}_N$ has a zero eigenvalue corresponding to conservation of crystal momentum; 
in three dimensions it is generally a nonzero matrix.
However, two dimensions are special. Consider, for example, the circular Fermi surface shown in Fig.~\ref{Fig:Scatt}(a). 
For the given ${\bf p}$ and ${\bf p}'$, there are only two possible processes for which ${\bf p}$, ${\bf p}'$, ${\bf k}$, and ${\bf k}'$ are on the Fermi surface and ${\bf p} + {\bf k} = {\bf p}' + {\bf k}'$; one process is described by ${\bf k}=-{\bf p}$ and ${\bf k}'=-{\bf p}'$, the other by ${\bf k}={\bf p}'$ and ${\bf k}'={\bf p}$.  
Since $\psi_{L'}({\bf p}) - \psi_{L'}({\bf p}')  + \psi_{L'}({\bf k}) - \psi_{L'}({\bf k}') =0$ in Eq.~(\ref{element_C}) for the both processes, $\hat{C}_N$ is a zero matrix. 
As discussed in \cite{MFII}, this result holds broadly for noncircular Fermi surfaces on a two-dimensional lattice, where ${\hat C}={\hat C}_U$.

Let us consider in more detail $\hat{C}$ for a two-dimensional circular Fermi surface with a Fermi wavenumber $k_F$. As shown in Fig.~\ref{Fig:Scatt}, for the given ${\bf p}$, ${\bf p}'$, and ${\bf G}$ (we include normal processes by ${\bf G} = {\bf 0}$), the two possible sets of solutions satisfying ${\bf p} - {\bf p}' + {\bf k} - {\bf k}' = {\bf G}$ for ${\bf p}$, ${\bf p}'$, ${\bf k}$, and ${\bf k}'$ on the Fermi surface is given by $({\bf k},{\bf k}')= ({\bf k}_\pm,{\bf k}_\pm')$. 
These solutions are explicitly given by
\be
k_{\pm,x}' &=& - k_{\mp,x} =  \frac{q_x}{2} \pm \frac{q_y}{2}\sqrt{\frac{4k_F^2}{q^2} -1},
\\
k_{\pm,y}' &=& - k_{\mp,y} =  \frac{q_y}{2} \mp \frac{q_x}{2}\sqrt{\frac{4k_F^2}{q^2} -1},
\ee
where ${\bf q} = {\bf p} - {\bf p}' - {\bf G}$. 
Let $\theta$ and $\theta'$ be the angles of ${\bf p}$ and ${\bf p}'$, respectively. 
Then, depending on the reciprocal vector ${\bf G} = (G_x, G_y)$, the angles $\alpha_{\pm}$ of ${\bf k}_\pm'$, which are functions of $\theta$ and $\theta'$, 
is obtained through
\be
\cos \alpha_{\pm}(\theta,\theta';{\bf G})
&=& \frac{1}{2} \left( \cos \theta - \cos \theta'- \frac{G_x}{k_F} \right)
\pm \frac{1}{2} \left( \sin \theta - \sin \theta'- \frac{G_y}{k_F} \right)
\nonumber\\&\quad&\times
\frac{\displaystyle{\sqrt{1-X^2(\theta,\theta';{\bf G})}}}{\displaystyle{X(\theta,\theta';{\bf G})}},
\\
\sin \alpha_{\pm}(\theta,\theta';{\bf G}) &=& \frac{1}{2} \left( \sin \theta - \sin \theta'- \frac{G_y}{k_F} \right) 
\mp \frac{1}{2} \left( \cos \theta - \cos \theta'- \frac{G_x}{k_F} \right) 
\nonumber\\&\quad&\times
\frac{\sqrt{\displaystyle{1-X^2(\theta,\theta';{\bf G})}}}{\displaystyle{X(\theta,\theta';{\bf G})}},
\ee
where
\be
X(\theta,\theta';{\bf G}) &=& \frac{1}{2} 
\left[ \left( \cos \theta - \cos \theta'- \frac{G_x}{k_F} \right)^2\right.
\left.+ \left( \sin \theta - \sin \theta'- \frac{G_y}{k_F} \right)^2 \right]^{1/2}.
\ee
The average over the Fermi surface is given by
\be
\frac{1}{N(0)} \int \frac{d {\bf p}}{(2\pi)^{2}} \delta({\tilde{\epsilon}}_{{\bf p}})
= \int_{-\pi}^{\pi} \frac{d\theta}{2\pi},
\ee 
where ${\tilde{\epsilon}}_{{\bf p}} = (p^2 - k_F^2)/2{\tilde m}$ and $N(0) = {\tilde m}/2\pi$, but the effect of the mass ${\tilde m}$ is canceled in $\hat{C}$. The Fermi-surface harmonics $\psi_L({\bf p})$ are given by 
\be
\psi_L({\bf p}) = \sqrt{2} \cos (2 L -1) \theta, 
\ee
where the $L=1$ Fermi surface harmonics is proportional to the $x$ component of the velocity or the momentum, $\psi_1({\bf p}) = \sqrt{2} \cos \theta = \sqrt{2}p_x/k_F$.
Therefore we obtain the matrix elements of $\hat{C}$ for the two-dimensional circular Fermi surface as
\be
C_{LL'} &=& 2 \sum_{{\bf G}\neq {\bf 0}} \int_{-\pi}^{\pi} \frac{d \theta d \theta'}{(2\pi)^{2}} 
w(\theta,\theta'; {\bf G}) 
\cos \{ (2L-1) \theta \} 
\nonumber
\\
&\quad& \times
\left[ \cos \{ (2L'-1) \theta \} - \cos \{ (2L'-1) \theta' \} \right.
\nonumber
\\
&\quad& 
\left. - \cos \{ (2L'-1) \alpha_{+}(\theta,\theta';{\bf G}) \} 
- \cos \{ (2L'-1) \alpha_{-}(\theta,\theta';{\bf G}) \} \right] .
\label{A_2DCFS}
\ee
Here $w(\theta,\theta';{\bf G})$ is a weight function satisfying 
$\sum_{{\bf G}} w(\theta,\theta';{\bf G})  = 1$,
\begin{align}
w(\theta,\theta';{\bf G})  &= \frac{\Theta\left(1 - X^2(\theta,\theta';{\bf G})  \right)}{
X(\theta,\theta';{\bf G})\sqrt{1 - X^2(\theta,\theta';{\bf G})}
}
\nonumber
\\
&\quad \times \left(
\sum_{{\bf G}'} 
\frac{\Theta\left(1 - X^2(\theta,\theta';{\bf G}')  \right)}{
X(\theta,\theta';{\bf G}')\sqrt{1 - X^2(\theta,\theta';{\bf G}')}
}
\right)^{-1} ,
\label{w_G}
\end{align}
where $\Theta(x)$ is the Heaviside step function. 
Note that ${\bf G}={\bf 0}$ is excluded from the summation for the reciprocal lattice vector ${\bf G}$ in Eq.~(\ref{A_2DCFS}), corresponding to the two-dimensional speciality ${\hat C}={\hat C}_U$ mentioned above. 
However, the sum over ${\bf G}'$ in Eq.~(\ref{w_G}) includes ${\bf G}'={\bf 0}$.

If $k_F {a_L} < \pi/2$, $w(\theta,\theta';{\bf G})$ vanishes for ${\bf G}\neq {\bf 0}$ [note that for example, $G_x/k_F >4$ for ${\bf G} = (2\pi/{a_L},0)$ and the step function in Eq.~(\ref{w_G}) is zero]. 
Then, by Eq.~(\ref{A_2DCFS}), $\hat{C}$ gets equal to a zero matrix, $\hat{C}  = \hat{0}$. Hence, the coefficient of the $T$ linear term in the electrical resistivity vanishes in the absence of Umklapp scattering.

The numerical evaluation of $C_{LL'}$ for various choices of $L_{max}$ has been carried out only for a circular Fermi surface of various radii on a square. The results for the Umklapp factor $\rho/\rho_{\rm SE}$ are presented in Sec. IV-C.

\section{Calculation of Seebeck coefficient} 

The basic formulas for Seebeck coefficient are given in Sec. IV B. We present here the results for its detailed derivation. The conclusions from the results, of importance to the experiments, are given in the main text.

For the critical fluctuations and their coupling to fermions given by Eq.~(\ref{assumption_fluctuation}), $Y_1(t)$ 
in Eq.~(\ref{L12general}) can be obtained as 
\be
Y_1(t)
&=& \frac{(-e)}{2\bar{g}^2N(0)T} \frac{{\tilde v}_{{\bf p}x}}{f(t)} \frac{1}{2}\,{\rm sech}\frac{t}{2}
 - \frac{(-e)\langle {\tilde v}_x^2 \rangle^{1/2}}{2\bar{g}^2N(0)T} 
\sum_{n=1}^{\infty}\varphi_{n,+}(t) \alpha_{n,1}
\nonumber \\
& & \times \frac{2 }{N(0)} \int \frac{d{\bf q}}{(2 \pi)^{d}} \delta({\tilde{\epsilon}}_{{\bf p}-{\bf q}}) 
\sum_{L} \bar{\psi}_L({\bf q}) \left[ \frac{1}{\lambda_{n,+} \hat{1} - \hat{B}} \right]_{L1}, 
\label{Y1LCO}
\ee
where we have defined
\be
\alpha_{n,1} &\equiv& \int_{-\infty}^{\infty} \varphi_{n,+}(x) \frac{1}{2} \,{\rm sech}\frac{x}{2} dx, 
\\
\bar{\psi}_L({\bf q}) &\equiv&
\frac{\displaystyle{
\sum_{\bf G}\int d{\bf k} \, \psi_L({\bf k}) \delta({\tilde{\epsilon}}_{\bf k}) 
\delta({\tilde{\epsilon}}_{{\bf k}+{\bf q}-{\bf G}})}
}{\displaystyle{\sum_{{\bf G}'}\int d{\bf k}' \, \delta({\tilde{\epsilon}}_{{\bf k}'}) 
\delta({\tilde{\epsilon}}_{{\bf k}'+{\bf q}-{\bf G}'})}} .
\ee
Note that $Y_1(t)$ is a function of not only the dimensionless energy variable $t$ but also the momentum ${\bf p}$, which is not restricted on the Fermi surface. For an ${\bf p}$ on the Fermi surface, 
$Y_1(t)$ can be expanded in the Fermi surface harmonics as $Y_1(t) = \sum_L \langle \psi_L |Y_1(t) \rangle \psi_L({\bf p})$ with
\be
\langle \psi_L |Y_1(t) \rangle &=& 
\frac{(-e)\langle {\tilde v}_x^2 \rangle^{1/2}}{2\bar{g}^2N(0)T} 
\sum_{n=1}^{\infty}\varphi_{n,+}(t) \alpha_{n,1}
\left[ \frac{\lambda_{n,+}}{\lambda_{n,+} \hat{1} - \hat{B}} \right]_{L1}.
\label{Y1Fermi}
\ee
Since the dc-electrical conductivity is given by $L_{11} =  2 \int_{-\infty}^{\infty} dt \sum_L \langle X_1(t) |\psi_L \rangle 
\langle \psi_L |Y_1(t) \rangle$ with $\langle X_1(t) |\psi_L \rangle = (-e) \langle {\tilde v}_x^2 \rangle^{1/2} \alpha_{n,1}$, 
Eq.~(\ref{Y1Fermi}) immediately leads to Eq.~(\ref{sigma_LCO}). 
As seen in Sec. IV C, for a large Fermi surface, the matrix ${\hat B}$ becomes small because of the frequent occurrence of Umklapp scattering. 
Note that substituting ${\hat B} = {\hat 0}$ into Eq.~(\ref{Y1Fermi}) and using $\sum_{n=1}^{\infty}\varphi_{n,+}(t) \alpha_{n,1}=(1/2)\,{\rm sech}(x/2)/f(t)$ reproduce the first term in Eq.~(\ref{Y1LCO}) for any ${\bf p}$ on the Fermi surface. Similarly, for such a large Fermi surface, the second term (the vertex correction) is much smaller than the first term (the self-energy correction) in Eq.~(\ref{Y1LCO}) for the ${\bf p}$ away from the Fermi surface. However, for a small Fermi surface without Umklapp scattering, the second term is important and leads to a divergent electrical conductivity. On the other hand, the vertex corrections to the thermal current vanish regardless of the size of the Fermi surface, and then $Y_2(t)$ is simply given by the self-energy correction,
\be
Y_2(t)
&=& \frac{1}{2\bar{g}^2N(0)} \frac{{\tilde v}_{{\bf p}x}}{f(t)} \frac{t}{2}\,{\rm sech}\frac{t}{2},
\label{Y2LCO}
\ee
which always leads to the finite thermal conductivity.

The off-diagonal transport coefficient can be obtained by substituting Eqs.~(\ref{Y1LCO}) and ~(\ref{Y2LCO}) into Eq.~(\ref{L12general}). Since the spectral function for the MFL can be approximated by 
$A({\bf p},\varepsilon) = {\tilde A}_{\bf p}(a^{-1}\varepsilon)$ with
\be
{\tilde A}_{\bf p} (x) = \frac{1}{\pi} \frac{{\tilde \Gamma}}{(x-\tilde{\epsilon}_{\bf p})^2 + {\tilde \Gamma}^2},
\ee
where $a^{-1} \propto |\ln T|$ and ${\tilde \Gamma} \propto T$ are independent of momentum, 
$A_0 = {\tilde A}_{\bf p} (0) \to \delta(\tilde{\epsilon}_{\bf p})$ for $T \to 0$ and the next order term can be written formally as $A_1(t) = a^{-1}Tt \,{\tilde A}_{\bf p}^{\,\prime} (0)$. Then we can derive ${\cal I}_1(t,t')$ by applying this formal expansion to the spectral functions in Eq.~(\ref{calI1}). As a result, the inverse of the single-particle renormalization factor $a$ can be taken out, and the off-diagonal transport coefficient is given by
\be
L_{12} &=& a^{-1} {\tilde L}_{12}.
\ee
Here ${\tilde L}_{12}$ does not include $a$, and some manipulations lead to
\be
{\tilde L}_{12} &=& - \frac{e T}{\bar{g}^2} 
\frac{1}{N(0)}\int \frac{d{\bf p}}{(2 \pi)^{d}} {\tilde v}_{{\bf p}x}^2 
\,\delta'(- \tilde{\epsilon}_{\bf p}) 
\nonumber\\&\quad&
+ \frac{e T}{\bar{g}^2} \frac{N'(0)}{N(0)} 
\frac{ \langle {\tilde v}_x^2 \rangle}{2}
\sum_{n=1}^{\infty}\alpha_{n,2}\alpha_{n,1}
\left[ \frac{\lambda_{n,+} }{\lambda_{n,+} \hat{1} - \hat{B}} \right]_{11}
\nonumber\\&\quad& 
+ \frac{2 e T}{{\bar{g}^2}} 
\frac{2 \langle {\tilde v}_x^2 \rangle^{1/2}}{N(0)^2} \sum_{L} 
\int \frac{d{\bf q}d{\bf p}}{(2 \pi)^{2d}}  
{\tilde v}_{{\bf p}x} 
\nonumber\\&\quad&\times
\,\delta(\tilde{\epsilon}_{\bf p}) \tilde{A}_{{\bf p}-{\bf q}}^{\,\prime}(0) 
\bar{\psi}_L({\bf q})
\left[ \frac{1}{\hat{1} - \hat{B}} \right]_{L1}
\nonumber
\\
&\quad& 
+ \frac{e T}{\bar{g}^2}
\frac{2\langle {\tilde v}_x^2 \rangle^{1/2} }{N(0)^2} 
\int \frac{d{\bf q}d{\bf p}}{(2 \pi)^{2d}} 
{\tilde v}_{{\bf p}x} 
\nonumber\\&\quad&\times
\left[\, \tilde{A}_{{\bf p}}^{\,\prime}(0)  \delta({\tilde{\epsilon}}_{{\bf p}-{\bf q}}) 
-\delta(\tilde{\epsilon}_{\bf p}) \tilde{A}_{{\bf p}-{\bf q}}^{\,\prime}(0)  \right]
\nonumber
\\
&\quad&\times
\sum_{n=1}^{\infty} \alpha_{n,2}\alpha_{n,1}\sum_{L} \bar{\psi}_L({\bf q}) \left[ \frac{1}{\lambda_{n,+} \hat{1} - \hat{B}} \right]_{L1} ,
\label{tilde L21}
\ee
where 
\be
\alpha_{n,2} \equiv \int_{-\infty}^{\infty} \varphi_{n,+}(t) 
\frac{t^2}{2}\,{\rm sech}\frac{t}{2} dt .
\ee
Now we extend the matrix ${\hat B}$ which represents the vertex corrections to the electric current as
\be
\left[{\hat B} (x,y)\right]_{LL'} &=& - \frac{2}{N(0)^2} \int \frac{d{\bf q}d{\bf p}}{(2 \pi)^{2d}}  
\psi_L({\bf p}) 
\nonumber\\&\quad&\times
\tilde{A}_{{\bf p}}(x) \tilde{A}_{{\bf p}-{\bf q}}(y) \bar{\psi}_{L'}({\bf q}) .
\ee
Then ${\hat B} = {\hat B}(0,0)$ for $T \to 0$. 
Introducing the partial derivatives of ${\hat B} (x,y)$ at $x=0$ and $y=0$ as
${\hat B}_x \equiv \partial {\hat B}(x,0)/\partial x|_{x=0}$ and 
${\hat B}_y \equiv \partial {\hat B}(0,y)/\partial y|_{y=0}$, 
we can write Eq.~(\ref{tilde L21}) as a rather compact form,
\begin{widetext}
\be
{\tilde L}_{12} &=& 
-  \frac{e \langle {\tilde v}_x^2 \rangle T}{\bar{g}^2}
\left( \frac{\langle {\tilde m}_{xx}^{-1} \rangle}{\langle {\tilde v}_x^2 \rangle} 
- \frac{1}{2}\frac{N'(0)}{N(0)} \right) 
- \frac{2 e \langle {\tilde v}_x^2 \rangle T}{{\bar{g}^2}}  
\left[{\hat B}_y (\hat{1} - \hat{B})^{-1} \right]_{11}
\nonumber\\ &\quad&
+ \frac{e \langle {\tilde v}_x^2 \rangle T}{\bar{g}^2}
\sum_{n=1}^{\infty} \alpha_{n,2}\alpha_{n,1}
\left[\left(\frac{1}{2}\frac{N'(0)}{N(0)}\hat{B} - {\hat B}_x + {\hat B}_y \right) (\lambda_{n,+} \hat{1} - \hat{B})^{-1} \right]_{11},\qquad
\label{tilde L21_v2}
\ee
\end{widetext}
where the average of the inverse mass tenser over the Fermi surface is given by
\be
\langle {\tilde m}_{xx}^{-1} \rangle = \frac{1}{N(0)}\int \frac{d{\bf p}}{(2 \pi)^{d}}
\frac{\partial {\tilde v}_{{\bf p}x}}{\partial p_x} 
\,\delta (\tilde{\epsilon}_{\bf p}) .
\ee

\section{Further details of velocity distribution and the Landau-Boltzmann transport equations} 

Noting Eq. (\ref{DRDA}),
we can rewrite Eq. (\ref{calI1}) as
\be
{\cal I}(p,p') &=& \pi W (p\!-\!p') 
\int_{p_1}\,
A (p_1) A (p_1\!-\!p\!+\!p') \,
\nonumber\\&\quad&\times
{\rm sech} \frac{\varepsilon_1}{2T} \, {\rm sech} \frac{\varepsilon_1\!-\!\varepsilon\!+\!\varepsilon'}{2T} 
\nonumber \\
&& + \pi \int_{p_1}\,
W (p\!-\!p_1)  A (p_1) A (p_1\!-\!p\!+\!p') 
\nonumber\\&\quad&\times
{\rm sech} \frac{\varepsilon_1}{2T} \,  {\rm sech} \frac{\varepsilon_1\!-\!\varepsilon\!+\!\varepsilon'}{2T}  
\nonumber \\
& & - \pi \int_{p_1}\,
W (p\!-\!p_1)  A (p_1) A (p\!+\!p'\!-\!p_1) 
\nonumber\\&\quad&\times
{\rm sech} \frac{\varepsilon_1}{2T} \,  {\rm sech} \frac{\varepsilon\!+\!\varepsilon'\!-\!\varepsilon_1}{2T},
\ee
where
\be
W(q)=\frac{|g({\bf q})|^4 {\rm Im} D_R (q)}{{\rm Im} \Pi_R (q)}.
\label{W_def}
\ee
Then Eq.~(\ref{Lambda3}) with $\omega = 0$ can be written as
\be
{\tilde v}_x (p) \,{\rm sech}\frac{\varepsilon}{2 T}
&=&  \pi 
\int_{p'}\!\int_{p_1}\,
W (p-p') A (p_1)A (p-p'+p_1)A(p')  
\nonumber \\
&\quad& \times
\big[\Phi (p) - \Phi (p') + \Phi (p_1)  - \Phi (p-p'+p_1)  \big]
\nonumber \\
&\quad& \times \,{\rm sech}\frac{\varepsilon_1}{2 T} 
\, {\rm sech} \frac{\varepsilon- \varepsilon' + \varepsilon_1}{2 T}
\,  {\rm sech} \frac{\varepsilon'}{2 T} .
\label{BS equation}
\ee
Assuming Landau's quasiparticles, we can show that Eqs.~(\ref{Kubo-1}) and (\ref{Kubo-Phi}) are equivalent, and Eq.~(\ref{BS equation}) corresponds to the Boltzmann equation for the problem of transport for fermions coupled to collective fluctuations in the limit of vanishing external frequency and wave vector. This is  shown below. In the main text, we have proceeded without such an assumption.

For Landau Fermi liquids or MFL, the single-particle spectral function is approximated by    
\begin{align}
A ({\bf p}, \varepsilon) \approx  a_{\bf p} \delta ( \varepsilon - \epsilon_{\bf p}^* ),
\label{marginal FL}
\end{align}
where $\epsilon_{\bf p}^*=a_{\bf p}{\tilde \epsilon}_{\bf p}$ is the energy of the quasiparticle. Even though in MFL the quasiparticle weight $a_{\bf p}$ is nonzero only because of a logarithmic factor, a Fermi surface is well defined so that for low temperatures we can use Eq. (\ref{marginal FL}) without introducing errors, where all the effects of $a_{\bf p}$ are canceled out in the theory such that $A({\bf p},0) = a_{\bf p} \delta(\epsilon_{\bf p}^*) = \delta({\tilde{\epsilon}}_{\bf p})$. [It is important to note that this cancellation occurs also for the MFL, where $a_{\bf p}(\epsilon_{\bf p}^*) \to 0$, as $|1/\ln(\epsilon_{\bf p}^*)|$.] Hence, the low-temperature conductivity can be obtained without using the Fermi liquid assumption, Eq.~(\ref{marginal FL}), and the results given by Eqs.~(\ref{sigma_FSH}) and (\ref{kappa_memory}) do not include any quasiparticle weight. However, Landau theory gives a familiar physical picture of quasiparticles carrying electric and thermal currents, so here we derive the Landau-Boltzmann transport equations for a model of fermions on a lattice scattering with collective fluctuations. 

Substituting Eq.~(\ref{marginal FL}) into Eq.~(\ref{BS equation}), we get
\be
v_{{\bf p}x}^*\,{\rm sech}\frac{\epsilon_{\bf p}^*}{2 T}
&=&  \frac{1}{2} 
 \int \frac{d {\bf p}' d {\bf k} d {\bf k}'}{(2\pi)^{2d}} 
S({\bf p},{\bf k};{\bf p}',{\bf k}') \,{\rm sech}\frac{\epsilon_{{\bf p}'}^*}{2 T} 
\, {\rm sech} \frac{\epsilon_{\bf k}^*}{2 T}
\,  {\rm sech} \frac{\epsilon_{{\bf k}'}^*}{2 T}
\nonumber \\
&\quad& \times 
\big[\Phi ({\bf p}) - \Phi ({\bf p}') + \Phi ({\bf k})  - \Phi ({\bf k}')  \big] ,
\label{Boltzmann}
\ee
where $v_{{\bf p}x}^*=a_{\bf p}{\tilde v}_x ({\bf p},\epsilon_{\bf p}^*)$ and 
$\Phi ({\bf p}) = \Phi ({\bf p},\epsilon_{\bf p}^*)$, 
and
\be
S({\bf p},{\bf k};{\bf p}',{\bf k}') &=& 2\pi W ({\bf p}-{\bf p}', \epsilon_{\bf p}^* - \epsilon_{{\bf p}'}^*) 
a_{\bf p} a_{{\bf p}'} a_{\bf k} a_{{\bf k}'} 
\nonumber \\
&\quad& \times 
\Delta^d({\bf p} -{\bf p}' + {\bf k} - {\bf k}')
\delta(\epsilon_{\bf p}^* - \epsilon_{{\bf p}'}^* + \epsilon_{\bf k}^* - \epsilon_{{\bf k}'}^*).
\ee
Noting that
\be 
f(\varepsilon)
\bar{f}(\varepsilon') 
f(\varepsilon_1)
\bar{f}(\varepsilon-\varepsilon'+\varepsilon_1)
=
\frac{1}{16}{\rm sech}\frac{\varepsilon}{2 T} 
\,{\rm sech}\frac{\varepsilon'}{2 T} 
\, {\rm sech} \frac{\varepsilon_1}{2 T}
\,  {\rm sech} \frac{\varepsilon-\varepsilon'+\varepsilon_1}{2 T}, 
\ee
where $\bar{f}(\varepsilon) = 1 - f(\varepsilon)$, we can write Eq.~(\ref{Boltzmann}) in the familiar form of the Landau-Boltzmann transport equation for the velocity distribution function $\Phi ({\bf p})$ as 
\be
v_{{\bf p}x}^* \left( - \frac{\partial f(\epsilon_{\bf p}^*)}{\partial \epsilon_{\bf p}^*}  \right)
&=&  
\frac{2}{T}
 \int \frac{d {\bf p}' d {\bf k} d {\bf k}'}{(2\pi)^{2d}} 
S({\bf p},{\bf k};{\bf p}',{\bf k}') 
f(\epsilon_{\bf p}^*)
\bar{f}(\epsilon_{{\bf p}'}^*) 
f(\epsilon_{\bf k}^*)
\bar{f}(\epsilon_{{\bf k}'}^*)
\nonumber \\
&\quad& \times 
\big[\Phi ({\bf p}) - \Phi ({\bf p}') + \Phi ({\bf k})  - \Phi ({\bf k}')  \big] .
\label{Landau-Boltzmann}
\ee
[From Eq.~(\ref{originalLBtransport}), that for the thermal-current distribution function is given by replacing $v_{{\bf p}x}^*$ with $v_{{\bf p}x}^*\epsilon_{\bf p}^*$.] 
It should be noted, however, that for fermions scattering with collective fluctuations, 
$W$ in $S$ is given by Eq.~(\ref{W_def}).

By substituting Eq.~(\ref{marginal FL}) for Eq.~(\ref{Kubo-Phi}), 
Eq.~(\ref{Kubo-1}) is rederived as
\be
\sigma(\omega,T) = 2 e^2 \int \frac{d {\bf p}}{(2 \pi)^{d}} 
\left( - \frac{\partial f(\epsilon_{\bf p}^*)}{\partial \epsilon_{\bf p}^*} \right) 
v_{{\bf p}x}^* \Phi ({\bf p},\omega).
\ee
From  Eqs.~(\ref{Lambda3}) and (\ref{Landau-Boltzmann}), 
$\Phi({\bf p},\omega)$  satisfies
\be
v_{{\bf p}x}^* \left( - \frac{\partial f(\epsilon_{\bf p}^*)}{\partial \epsilon_{\bf p}^*}  \right)
&=&  
-i\omega 
\left( - \frac{\partial f(\epsilon_{\bf p}^*)}{\partial \epsilon_{\bf p}^*}  \right)
\left[ \Phi ({\bf p},\omega) - 
\int \frac{d {\bf p}'}{(2 \pi)^{d}} 
g({\bf p},{\bf p}') 
\left( - \frac{\partial f(\epsilon_{{\bf p}'}^*)}{\partial \epsilon_{{\bf p}'}^*}  \right) 
\Phi ({\bf p}',\omega)
\right]
\nonumber\\
&\quad&+
\frac{2}{T}
 \int \frac{d {\bf p}' d {\bf k} d {\bf k}'}{(2\pi)^{2d}} 
S({\bf p},{\bf k};{\bf p}',{\bf k}') 
f(\epsilon_{\bf p}^*)
\bar{f}(\epsilon_{{\bf p}'}^*) 
f(\epsilon_{\bf k}^*)
\bar{f}(\epsilon_{{\bf k}'}^*)
\nonumber\\&\quad& \times 
\big[\Phi ({\bf p},\omega) - \Phi ({\bf p}',\omega) + \Phi ({\bf k},\omega)  - \Phi ({\bf k}',\omega)  \big] ,
\ee
where $g({\bf p},{\bf p}') = a_{\bf p}\Gamma^k({\bf p},\epsilon_{\bf p}^*,{\bf p}', \epsilon_{{\bf p}'}^*)a_{{\bf p}'}$.

\bibliography{Maebashi_V_refs.bib}
\end{document}